\documentclass[11pt]{article}

\usepackage{graphicx}
\usepackage{amsmath}
\usepackage{dcolumn}
\usepackage{bm}
\usepackage{slashed}
\usepackage[bookmarks,bookmarksnumbered,colorlinks=true,anchorcolor=blue,
linkcolor=blue,urlcolor=blue,citecolor=blue,breaklinks=true]{hyperref}
\usepackage[utf8]{inputenc}
\usepackage{authblk}

\usepackage{color}
\usepackage{ulem}

\newcommand{\be}{\begin{equation}}
\newcommand{\ee}{\end{equation}}
\newcommand{\ba}{\begin{eqnarray}}
\newcommand{\ea}{\end{eqnarray}}
\def\bea{\begin{eqnarray}}
\def\eea{\end{eqnarray}}
\newcommand{\gsim}{\mathrel{\hbox{\rlap{\lower.55ex \hbox {$\sim$}}
                   \kern-.3em \raise.4ex \hbox{$>$}}}}
\newcommand{\lsim}{\mathrel{\hbox{\rlap{\lower.55ex \hbox {$\sim$}}
                   \kern-.3em \raise.4ex \hbox{$<$}}}}

\newcommand{\bal}{\begin{aligned}}
\newcommand{\eal}{\end{aligned}}

\def\roughly#1{\mathrel{\raise.3ex\hbox{$#1$\kern-.75em%
\lower1ex\hbox{$\sim$}}}}
\def\lsim{\roughly<}
\def\gsim{\roughly>}

\def\({\left(}
\def\){\right)}
\def\[{\left[}
\def\]{\right]}
\def\<{\langle}
\def\>{\rangle}

\def\a{{\alpha}}

\def\m{{\mu}}
\def\n{{\nu}}

\usepackage{xcolor}

\title{\bf {Holographic renormalized Entanglement and entropic $c-$function}}

\author[1]{Mitsutoshi Fujita \thanks{fujitamitsutoshi@usc.edu.cn}}
\author[2,3]{Song He\thanks{hesong@jlu.edu.cn }}
\author[4]{Yuan Sun\thanks{sunyuan@csu.edu.cn}}
\author[5]{Jun Zhang\thanks{jzhang163@crimson.ua.edu}}
\affil[1]{School of Nuclear Science and Technology, University of South China, Hengyang 421001, China}
\affil[2]{Center for Theoretical Physics and College of Physics, Jilin University, Changchun 130012, People’s Republic of China}
\affil[3]{Max Planck Institute for Gravitational Physics (Albert Einstein Institute), Am M\"uhlenberg 1, 14476 Golm, Germany}
\affil[4]{School of Physics and Electronics, Central South University, Changsha 418003, China}
\affil[5]{Department of Physics and Astronomy, University of Alabama, 514 University Boulevard, Tuscaloosa, AL, 35487, USA}

\date{}

\begin{document}

\maketitle
%\vspace{0.1in}

\begin{abstract}
We compute holographic entanglement entropy (EE) and the renormalized EE in AdS solitons with gauge potential for various dimensions. The renormalized EE is a cutoff-independent universal component of EE. Via Kaluza-Klein compactification of $S^1$ and considering the low-energy regime, we deduce the $(d-1)$-dimensional renormalized EE from the odd-dimensional counterpart. This corresponds to the shrinking circle of AdS solitons, probed at large $l$. The minimal surface transitions from disk to cylinder dominance as $l$ increases. The quantum phase transition occurs at a critical subregion size, with renormalized EE showing non-monotonic behavior around this size. Across dimensions, massive modes decouple at lower energy, while degrees of freedom with Wilson lines contribute at smaller energy scales.

\end{abstract}

\newpage

\allowdisplaybreaks

\flushbottom

\section{Introduction}
Quantum entanglement entropy stands as a pivotal concept in quantum mechanics, offering insight into the level of entanglement among distinct segments of a quantum system. By quantifying the entanglement between different components, this entropy provides a metric for the extent of shared information. Its implications span various facets of quantum mechanics, encompassing quantum information theory, black hole physics, and condensed matter physics. Notably, the entanglement entropy of subsystem A quantifies the entangled degrees of freedom within a given quantum field theory~\cite{Cardy, Review}. Within the context of condensed matter physics, this entropy displays divergence at critical junctures of quantum critical phase transitions, assuming the role of an order parameter~\cite{Vidal:2002rm}. This phenomenon encapsulates the geometric essence of field theories, manifested in an area law that draws parallels between subregion entanglement entropy and black hole entropy.

Introducing the Ryu-Takayanagi formula establishes a holographic counterpart for entanglement entropy~\cite{Ryu:2006bv, Ryu:2006ef, Nishioka:2009un}, emerging as a robust tool for dissecting strongly coupled systems traditionally resistant to conventional analysis. In specific contexts, this formula has served as an order parameter, signaling the onset of confinement/deconfinement phase transitions within confining gauge theories~\cite{Nishioka:2006gr, Klebanov:2007ws, Buividovich:2008gq, Dudal:2016joz, Dudal:2018ztm, Mahapatra:2019uql}.~\footnote{On the other hand, holographic quark anti-quark potential can distinguish confinement and topological phases~\cite{Jokela:2020wgs}.} The transitions emanate from the interplay between two minimal surfaces, resulting in the post-transition confinement phase entanglement entropy becoming trivial at the infrared limit. Moreover, the holographic entanglement entropy (HEE) emerges as a probing tool for phase transitions in holographic superconductors~\cite{Albash:2012pd, Cai:2012sk, Cai:2012nm, Arias:2012py, Kuang:2014kha, Zangeneh:2017tub, Das:2017gjy} as well as for unveiling topological phases of matter~\cite{Baggioli:2023ynu}.

The entropic $c-$function concept provides deeper insights into entanglement entropy in quantum systems \cite{Casini:2004bw, Casini:2006es}. It represents the logarithmic derivative of entanglement entropy with respect to subsystem size, revealing the intricate interplay between entanglement and subsystem dimensions. The general entropic $c-$function, proposed by \cite{Nishioka:2006gr}, efficiently quantifies degrees of freedom in confining theories and yields the central charge of the corresponding conformal field theory (CFT) \cite{Myers:2012ed}. In the context of a quantum field theory dual to an AdS soliton, the behavior of the entropic $c-$function, as it decreases with increasing length, effectively serves as a probe for the deconfinement phase transition. Recently, a study \cite{Fujita:2020qvp} computed the entropic $c-$function for a striped entangling surface in the same background. Intriguingly, this function displays non-monotonic behavior with increasing background gauge field strength. Importantly, the entropic $c-$function for the strip incorporates both A-type and B-type anomalies due to the coexistence of these two anomaly types. \cite{Ghasemi:2019xrl} has derived constraints on anisotropic RG flows from holographic entanglement entropy. 

Our focus lies in assessing the degrees of freedom through entanglement entropy using a spherical entangling surface, with a specific emphasis on anomaly effects. The renormalized entanglement entropy, calculated from the entanglement entropy of this spherical surface, offers a solution to the issue of cutoff dependence \cite{Liu:2012eea}. This renormalized quantity, independent of the cutoff, measures degrees of freedom in quantum entangled states at an energy scale of $1/l$, yielding the central charge in conformal field theory (CFT). In the context of four-dimensional (4D) CFT, it manifests as the A-type anomaly, in alignment with the C-theorem: the renormalized entanglement entropy decreases in the infrared (IR) limit as anticipated. \cite{Ghasemi:2017pke} derived the renormalized entanglement entropy for a kink region which reduced to a universal positive finite term in the UV limit.

The computation of renormalized entanglement entropy for a spherical entangling surface within the AdS soliton framework with a gauge potential remains unexplored. The gauge potential's interpretation in this background involves a twisted boundary condition along a circle within the cigar direction. This contributes to the negative Casimir energy of the dual field theory, which can eventually become positive. An intriguing aspect emerges from the interplay of Wilson lines, capable of inducing mass shifts in charged particles \cite{Polchinski:1998rq}. It becomes desirable to capture such alterations through the renormalized entanglement entropy. Conversely, contrasting the analysis presented in \cite{Fujita:2020qvp} for a striped entangling surface in the same background, the renormalized entanglement entropy in $R^{1,2}\times S^1$ quantum field theory exhibits solely B-type anomaly characteristics. Generally, this quantity doesn't adhere to the C-theorem. Therefore, an engaging pursuit lies in investigating this aspect, including scenarios in higher-dimensional cases.

This study focuses on introducing a holographic renormalized entanglement entropy (HREE), which encompasses the finite portion of entanglement entropy. We extensively investigate HREE's behavior across diverse scenarios to uncover the universal properties of quantum phase transitions. Our analysis reveals a phase transition between disk and cylinder geometries. The dominance of the disk type is evident for small $l$, while the cylinder shape prevails for larger $l$. This critical size marks the occurrence of a quantum phase transition. Notably, this transition is an outcome of the large $N$ limit and is absent in free theory. 
The anticipated function of the renormalized entanglement entropy is to quantify the degrees of freedom in the dual quantum field theory (QFT). We probe HREE's response to changes in operator mass and gauge potential. Specifically, as the operator mass decreases, we expect HREE to increase significantly for larger $l$, given the decoupling of massive degrees of freedom.

%***It is expected that the renormalized entanglement entropy can count the degrees of freedom of dual QFT. The renormalized entanglement entropy will be useful for analyzing changes of mass spectrum due to the gauge potential. 
%Dual QFT is confining theory. It contains glueball spectrum like , which has dependence on the gauge potential (Wilson lines).*** 

The rest of this paper is organized as follows. In section II, we analyze the UV structure of the entanglement entropy. We derive the renormalized entanglement entropy of QFT dual to AdS solitons with gauge potential. We discuss properties of the renormalized entanglement entropy in both odd and even dimensions. Section III mainly focuses on holographic stress-energy tensors in AdS solitons with the gauge potential. In sections IV and V, we analyze the quantum phase transition of HREE in higher dimensional backgrounds. We end with conclusions and prospects in section VI. Some calculation details are presented in the appendices.

\section{The UV structure of the entanglement entropy}
The $d+1$-dimensional AdS solitons with gauge potential exhibit a geometry akin to a cigar. In this setup, a compact circle gradually contracts to zero size in the bulk, completing the geometry. This behavior is detailed in \cite{Fujita:2020qvp}. The dual theory on $R^{1,d-2}\times S^1$ transforms into a confining theory with a discernible energy gap in this context. Notably, this theory incorporates Wilson lines along the $S^1$ direction, which inherently alters the boundary conditions.

We calculate the entanglement entropy in this theory. The spherical entangling surface is chosen to have the topology $S^{d-3}\times S^1$, where the entangling surface wraps another circle $S^1$~\cite{Ishihara:2012jg}. For $d=4$, $S^1\times S^1$ is a cylinder with one identified direction, i.e., a torus. Because a spherical entangling surface for QFT on $R^{1,d-1} $ has a different topology $S^{d-2}$, we find that the UV scaling structure is different from those of QFT on $R^{1,d-1}$. According to~\cite{Ishihara:2012jg}, the UV divergent structure of entanglement entropy $S_{UV}$ is of the form
\ba\label{SUV0}
S_{UV}=\dfrac{L_{\phi}}{R}S_{UV,0},
\ea
where $S_{UV,0}$ is the UV structure of entanglement entropy of QFT  $S_{EE}^{(0)}$ on $R^{1,d-1}$ and $L_{\phi}$ is the periodicity along a circle $S^1$ of the cigar.

Two UV scaling structures are related to each other. Using \eqref{SUV0} and operating differentiation on $S_{EE}$, the renormalized entanglement entropy (the UV-independent part of the entanglement entropy) then becomes \cite{Ishihara:2012jg}
\ba\label{SUV2}
&S_{ren}=\dfrac{1}{R}f_d(R\partial_R)RS_{EE}=L_d(R\partial_R )S_{EE} \nonumber \\
&=\begin{cases} &
\dfrac{1}{(d-2)!!}R\partial_R (R\partial_R -2)\dots (R\partial_R -(d-3))S_{EE},\quad d=\text{odd}, \\
& \dfrac{1}{(d-2)!!}(R\partial_R +1)(R\partial_R -1)\dots (R \partial_R -(d-3))S_{EE},\quad d=\text{even},
\end{cases}
\ea
where $f_d$ defines the operation of the renormalized entanglement entropy for $R^{1,d-1}$~\cite{Liu:2012eea}:
\ba\label{SUV3}
f_d(R\partial_R)S_{EE}^{(0)}=\begin{cases} &
\dfrac{1}{(d-2)!!}(R\partial_R-1) (R\partial_R -3)\dots (R\partial_R -(d-2))S_{EE}^{(0)},\quad d=\text{odd}, \\
& \dfrac{1}{(d-2)!!}R\partial_R (R\partial_R -2)\dots (R \partial_R -(d-2))S_{EE}^{(0)},\quad d=\text{even}.
\end{cases}
\ea
Recall that the first line in \eqref{SUV2} is equivalent to the $d-1$ dimensional renormalized entanglement entropy on $R^{1,d-2}$ (the second line in \eqref{SUV3}) up to coefficients. 

Especially we obtain 
\ba
&S_{ren}&=R\partial_R S_{EE} \quad \text{for $d=3$}, \label{SRE4} \\
&S_{ren}&=\dfrac{1}{3}R\partial_R (R\partial_R -2)S_{EE} \quad \text{for $d=5$}. \label{SRE5}
\ea
The formula \eqref{SRE4} for QFT on $R^{1,1}\times S^1$ corresponds to the well-established expression of the entropic $c$-function on $R^{1,1}$. Through Kaluza-Klein reduction along $S^1$, the renormalized entanglement entropy effectively embodies the 2-dimensional entropic $c-$function in the low-energy limit. In systems respecting Lorentz symmetry, the 2-dimensional entropic $c$-function is both non-negative and monotonically increasing. For $R\ll L_{\phi}$ (in the UV limit), the renormalized entanglement entropy mirrors the behavior of a 3-dimensional system. The subregion's topology is not a disk $D$ but rather $L\times S^1$ with an interval $L$, while the entangling surface forms $S^1$. Moving to formula \eqref{SRE5} for QFT on $R^{1,3}\times S^1$, it captures one variant of the 4-dimensional renormalized entanglement entropy on $R^{1,3}$. Kaluza-Klein reduction along $S^1$ approximates the renormalized entanglement entropy on $R^{1,3}$ in the low-energy regime. However, the behavior of the 4-dimensional renormalized entanglement entropy can be either negative or positive, displaying non-monotonic tendencies. In this scenario, the subregion's topology does not correspond to a ball $B^4$ but rather $B^3\times S^1$, while the entangling surface takes the form of $S^2\times S^1$.

%\textcolor{red}{To do something novel, we can check if the normalized entanglement entropy satisfies some inequality, like, strong subadditivity, etc..}

%3 dimensional renormalized entanglement entropy for QFT on $R^{1,2}$ has the same property as \eqref{SRE4}: it is also %non-negative and monotonically decreasing with $R$. 

\subsection{Renormalized entanglement entropy of 4d QFT}
This section examines the entanglement entropy and renormalized entanglement entropy for various cases: a free scalar, Dirac fermion, and a 4-dimensional conformal field theory (CFT) on $R^{1,2}\times S^1$. Additionally, we provide an overview of the trace anomaly in general CFT, which is intricately linked to the logarithmic term present in the entanglement entropy.

The entanglement entropy can be derived from the effective action $w=-\log Z$ in a $d(=4)$-dimensional manifold featuring conical singularities. By taking the limit $n\to 1$, the entanglement entropy assumes an analytical expression involving $w$ on a manifold with such singularities. Notably, the effective action $w$ generally exhibits logarithmic divergence, which is connected to the concept of conformal anomaly. We consider the infinitesimal rescaling $g^{\mu\nu}\to (1-2\delta\lambda)g^{\mu\nu}$. 
 We then have 
\ba\label{RES6}
\dfrac{dw}{d\lambda}=-2 g^{\mu\nu}\dfrac{\delta w}{\delta g^{\mu\nu}}=-\int d^4x \sqrt{g}\langle T^{\mu}{}_{\mu} \rangle
\ea
The equation stands as the trace anomaly, indicating the deviation from the traceless condition $\langle T^{\mu}{}_{\mu}\rangle =0$ within Quantum Field Theory (QFT) for Conformal Field Theory (CFT). The trace anomaly is characterized by polynomials of the curvature tensor, a formulation contingent on the dimension $d$. Notably, in odd dimensions, the trace anomaly must satisfy the condition of vanishing.

When we define the length scale $R_1$ of the subregion, this scale is related to rescaling the metric \eqref{RES6}.  Thus, one obtains the following formula:
\be\bal
R_1\partial_{R_1}S_A=&\lim_{n\to 1}(-2\partial_n\int d^{d+1}xg_{\mu\nu}(x)\dfrac{\delta}{\delta g_{\mu\nu}(x)}[w-nw|_{n=1}])   \\
&\dfrac{1}{2\pi}\lim_{n\to 1}\partial_n \Big(\langle \int d^{d+1}x\sqrt{g}T_{\mu}{}^{\mu}(x)\rangle_{M_n}-n\langle \int d^{d+1}x\sqrt{g}T_{\mu}{}^{\mu}(x)\rangle_{M_1}  \Big).
\eal\ee
Here, the entanglement entropy has been replaced with a function of the partition function in the manifold with conical singularities. The above-mentioned formula relates the entanglement entropy to the trace anomaly. 

To evaluate the entanglement entropy, we examine a subsystem $\sigma$ with a cylindrical configuration where one direction is identified as $\phi\sim \phi +L_{\phi}$. Interestingly, this subsystem aligns with the one in QFT corresponding to $d+1$-dimensional AdS solitons. Due to this, a conical singularity arises, characterized by a curvature tensor proportional to a delta function. The resulting logarithmic contribution to the entanglement entropy is expressed as $S_{EE}=s \log (\epsilon /R_1)+\dots $, where $\epsilon$ represents the ultraviolet cut-off. Remarkably, this logarithmic term can also be derived by integrating the entanglement entropy of a $3d$ free theory~\cite{Casini:2009sr, Huerta:2011qi}. $s$ is expressed in terms of extrinsic curvatures~\cite{Solodukhin:2008dh}. 
 According to~\cite{Casini:2009sr,Huerta:2011qi}, $s$ becomes
 \ba s=\dfrac{a}{180}\int_{\partial \sigma}d^2x\sqrt{h}E_2+\dfrac{c}{120}\int_{\partial \sigma}d^2x\sqrt{h}I_2,  \ea
where $E_2$ is the Euler density and $I_2$ is a Weyl invariant.  Compared with the normalization of~\cite{Liu:2012eea}, we have $a=360a_4$ and $c=120c_4$. $c=1$  for a real scalar and $c=6$ for Dirac {fermion}. Coefficients are consistent with trace anomalies.

For CFT on a cylinder of length $L_{\phi}$  and radius $l$, $s$ becomes
\ba
s=\dfrac{c}{240}\dfrac{L_{\phi}}{l}.
\ea

By using the renormalized entanglement entropy for cylinder type topology in \eqref{SUV2}, we obtain 
\ba
S_{ren}=\dfrac{1}{2}(l\partial_l +1)(l\partial_l -1)S_{EE}=s=\dfrac{cL_{\phi}}{240 l}.
\ea 
This formula shows that the renormalized entanglement entropy agrees with the coefficient $s$. Furthermore, according to~\cite{Liu:2012eea}, $s$ agrees with the renormalized entanglement entropy for a spherical entangling surface $S^2$ as follows: 
\ba
S_{ren}=\dfrac{1}{2}R\partial_R (R\partial_R -2)S_{EE}=s=\dfrac{a}{90}.
\ea

\section{The $AdS$ soliton with the gauge potential}
The AdS soliton is achieved through a double Wick rotation of the AdS black hole, following the Einstein equation. It corresponds to the QFT system with anti-periodic boundary conditions~\cite{Horowitz:1998ha}. In our investigation, we apply this approach to the Reissner Nordstrom $AdS$ black hole~\cite{Hartnoll:2009sz}, performing an analytical continuation of the metric in both temporal and spatial dimensions. The metric of the $AdS$ soliton with the gauge potential becomes \cite{Fujita:2020qvp}
\be\bal\label{METd1}
ds_{d+1}^2=& \dfrac{L^2}{z^2}\Big(\dfrac{dz^2}{f_d(z)}+f_d(z)d\phi^2-dt^2+dR^2+R^2d\Omega_{d-3}\Big),   \\
A_{\phi}=& a_{\phi}^{(0)}\Big(1-\Big(\dfrac{z}{z_+}\Big)^{d-2}\Big),
\eal\ee
where 
\ba
f_d(z)=1-\Big(1+\tilde{\epsilon}z_+^2a_{\phi}^2 \Big)\Big(\dfrac{z}{z_+}\Big)^d+\tilde{\epsilon}z_+^2a_{\phi}^2 \Big(\dfrac{z}{z_+}\Big)^{2(d-1)}.
\ea
Here, we set $\tilde{\epsilon}=-1$~\footnote{$\tilde{\epsilon}=1$ for the Reissner Nordstrom $AdS$ black hole.}, and define $a_{\phi}^2=a_{\phi}^{(0)2}/\gamma^2$, where $\gamma^2 =\frac{(d-1)g_e^2L^2}{(d-2)\kappa^2}$ is a dimensionless parameter. The gauge field $a_{\phi}^{(0)}$ acts as the source for the conserved current and induces a non-zero VEV for the current, $\langle J_{\phi} \rangle \neq 0$. Alternatively, this gauge field can be interpreted as a Wilson line, altering the boundary condition (twisted boundary condition) due to a gauge transformation. As the Wilson line vanishes at the tip of the soliton ($z=z_+$), the gauge connection remains regular there. The radial coordinate $z$ in \eqref{METd1} is confined to $z\le z_+$, while the $\phi$ direction follows the periodicity $\phi\to \phi +1/M_0$ to prevent conical singularities.
The Kaluza-Klein mass $M_0$ of the $\phi$ direction is given by 
\ba\label{KKM0}
M_0=\dfrac{1}{4\pi z_+}\Big(d+(d-2)\dfrac{a_{\phi}^2}{z_+^2}\Big).
\ea
The formula \eqref{KKM0} can also be rewritten in terms of $M_0$ and $a_{\phi}$ as follows:
\ba\label{ZP8}
z_{+}=\dfrac{d}{2\pi M_0\pm \sqrt{4\pi^2 M_0^2-d(d-2)a_{\phi}^2}}.
\ea 
There is also a minus branch. However, $z_+$ is divergent at small $a_{\phi}$ in that case, and the background does not approach the $AdS_{d+1}$ soliton. It can be shown that the solution with the plus sign in \eqref{ZP8} is always more stable than the one in the minus branch.

The boundary stress tensor $T^{(0)}_{\m\n}$ for field theory dual to the above background was computed in our previous work  \cite{Fujita:2020qvp}. Here, we quote the results of the boundary stress tensor for later use. For more details, please refer to  \cite{Fujita:2020qvp}
\be \label{Ttt2}
T^{(0)}_{tt}=-T^{(0)}_{x^ix^i}=-\frac{L^{d-1}}{2\kappa^2}\frac{1}{z^d_+}\Big(1-{z^2_+ a^2_\phi}\Big)=-\frac{L^{d-1}}{2\kappa^2}\frac{1}{z^d_+}\bar{\a}_\phi,
\ee
\be
T^{(0)}_{\phi\phi}=\frac{dL^{d-1}}{2\kappa^2}\frac{1}{z^d_+}\Big(1-{z^2_+ a^2_\phi}\Big)\Big(-1+\frac{1}{d}\Big).
\ee
with 
\be 
 \bar{a}_\phi =1-\Big({z_+ a_\phi}\Big)^2.
\ee
The boundary energy then is ~\cite{Balasubramanian:1999re}
\ba\label{Ttt1}
M= -\dfrac{V_{d-2}}{M_0}\frac{L^{d-1}\bar{a}_\phi}{2\kappa^2 z_+^d }.
\ea
The boundary energy can change the sign when we change Wilson lines (gauge potential). 
 \ba
 \begin{cases}
M<0 \quad \text{$z_+a_{\phi} <1$} \\
M>0  \quad \text{$z_+a_{\phi} >1$}.    
 \end{cases}
 \ea 
 In other words, $M$ is negative for $a_{\phi} <2\pi M_0/(d-1)$, while it can become positive when  $a_{\phi} >2\pi M_0/(d-1)$. For $a_{\phi}=0$, it realizes Casimir energy of $4d$ SYM theory on $R^3\times S^1$~\cite{Horowitz:1998ha}. This behavior is analogous to Casimir energy of fields with twisted boundary conditions in $2d$ CFT~\cite{Polchinski:1998rq}.  Casimir energy is different between the periodic and anti-periodic boundary conditions.

\section{The holographic entanglement entropy}
\begin{figure}[htbp]
     \begin{center}
\includegraphics[height=2.5cm,clip]{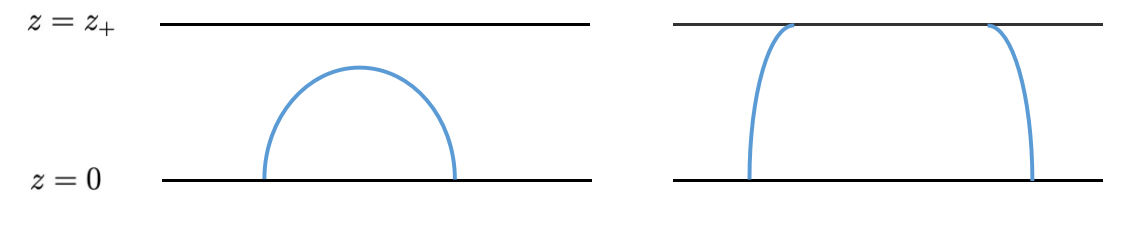} 
\caption{The RT surfaces corresponding to the small (left) and large (right) subsystem in the AdS-Soliton background. If the subsystem is a spherical region, then the RT surface has a disk (cylinder) topology for the small (large) subsystem.}\label{RTs2}
    \end{center}
\end{figure}

In this section, we compute the entanglement entropy~\cite{Ryu:2006bv, Ryu:2006ef}.  The entangling surface is specified by $z=0$ at $R=l$, and $0\le \phi \le L_{\phi}$  at a constant time slice in the background (\ref{METd1}). Its topology becomes $S^1\times S^{d-3}$, where $S^1$ and $S^{d-3}$  be of radius $L_\phi$ and $l$ respectively.  Note that the topology of the entangling surface differs from the theory without the compactification of the $\phi$ direction. The surface action becomes 
\ba\label{ACT3}
A=\int d^{d-1}x\mathcal{L}=\Omega_{d-3}L_{\phi}{L^{d-1}}\int dz \dfrac{R^{d-3}}{z^{d-1}}\sqrt{1+f \dot{R}^2},
\ea
where $\mathcal{L}=\sqrt{\det g_{ind}}$ and $g_{ind}$ is the induced metric. The holographic entanglement entropy is given by 
\ba
S_{EE}=\dfrac{2\pi}{\kappa^2}A
\ea
with   $A$   minimized. 
Recall that $L^{d-1}/\kappa^2$ is dimensionless. We omit the $AdS$ radius ($L=1$) for the convenience. 
We solve EOM derived from \eqref{ACT3} to obtain the minimal surface. 
The EOM of $R$ become~\footnote{EOM in terms of $z$ is
\ba
&-f(z) \left((d-1) R z'(R)^2+z(R) \left((d-3) z'(R)+r z''(R)\right)\right)+(1-d)R f(z)^2 \nonumber \\
&-(d-3) z(R) z'(R)^3=0.
\ea} 
\ba\label{REO}
&2 z \left(d-R(z) f'(z) R'(z)-3\right)+f(z) (2 (d-3) z R'(z)^2-R(z) (-2 (d-1) R'(z) \nonumber \\
&+z f'(z) R'(z)^3+2 z R''(z)))+2 (d-1) f(z)^2 R(z) R'(z)^3=0.
\ea
One should specify the IR boundary condition. In fact, there are two kinds of RT surfaces, as drawn in Fig.(\ref{RTs2}) schematically. The turning point of the disk type RT surface is $R(z_t)=0$. Moreover, the surface of the disk type is smooth at the bulk. The embedding scalar must satisfy $R'(z_t)=\infty$. The surface ends at the tip of the soliton $z_t=z_+$ for a cylinder case.  

Varying $A$ in terms of $l$ and fixing $z=\epsilon$, the Hamiltonian-Jacobi method (please refer to appendix \ref{HJmethod} for a brief review of this method)
becomes~\cite{Liu:2012eea}
\ba\label{HJ7}
\dfrac{dA}{dl}=-H(z_t)\dfrac{dz_t}{dl}-\Pi (\epsilon)\dfrac{dR(\epsilon )}{dl}=-\Pi (\epsilon)\dfrac{dR (\epsilon)}{dl},
\ea
where 
\ba
\Pi =\dfrac{\partial \mathcal{L}}{\partial \dot{R}}=\Omega_{d-3}{L^{d-1}}L_{\phi}\dfrac{R^{d-3}f\dot{R}}{z^{d-1}\sqrt{1+f\dot{R}^2}},\quad H=\Pi \dot{R}-\mathcal{L}=-\dfrac{\Omega_{d-3}L_{\phi}R^{d-3}{L^{d-1}}}{z^{d-1}\sqrt{1+f\dot{R}^2}}.
\ea
The first term of \eqref{HJ7} drops out due to the following IR boundary conditions
\be\bal
&R(z_t)=0, \quad \dot{R}(z_t)=\infty , \quad H(z_t)=0, \quad \text{for a disk}, \nonumber \\
&\dfrac{dz_t}{dl}=\dfrac{d\epsilon }{dl}=0, \quad \text{for a cylinder}.
\eal\ee
Equation \eqref{HJ7} only depends on the solution near the $AdS$ boundary, and an asymptotic expansion is useful. 

We compute the asymptotic expansion of the embedding scalar near $z=0$. The UV behavior of $R(z)$ has the following ansatz:
\ba\label{REP11}
R(z) =l+b_0\log \dfrac{z}{l}+\sum_{n=1}\Big(a_n+b_n\log \Big(\dfrac{z}{l}\Big)\Big)z^n,
\ea
where the log term arises in \eqref{REP11} similar to the Fefferman-Graham expansion of fields in the $AdS$ spacetime~\cite{Henningson:1998gx,de Haro:2000xn}. We can determine coefficients $a_n$ and $b_n$ after substituting the ansatz mentioned above into \eqref{REO}.
Below we will analyze the cases with $d=4,5,3$ in detail.

%\subsection{***}

\subsection{$d=4$}
Let us begin with the $d=4$ case, where the boundary QFT lives on $R^{1,2}\times S^1$, and the topology of the entangling surface becomes $S^1\times S^1$. By substituting expansion of $R(z)$ near the boundary $z=0$ (\ref{REP11}) into equation of motion, one can obtain 
\ba\label{EXP}
R(z)=l-\dfrac{z^2}{4l}+a_4(l) z^4+\dfrac{z^4}{32 l^3}\log \dfrac{z}{l}+\dots  
\ea
Here the higher order terms can be determined by parameters $l$ and $a_4(l)$. The coefficient $a_4(l)$ can not be determined from the UV expansion of the EOM. Instead, $a_4(l)$ has information determined by the IR boundary condition. 

Substituting \eqref{EXP} into \eqref{HJ7}, the $l$ derivative of the surface becomes
\ba\label{dAl21}
\dfrac{dA}{2\pi L_{\phi}dl}=-4l a_4(l)-\dfrac{3}{32l^2}+\dfrac{1}{2\epsilon^2}-\dfrac{1}{8l^2}\log \Big(\dfrac{\epsilon}{l}\Big)+\dots
\ea
The divergent structure of \eqref{dAl21} is
\ba
\dfrac{1}{2\epsilon^2}-\dfrac{1}{8l^2}\log \Big(\dfrac{\epsilon}{l}\Big).
\ea
Divergent pieces are defined up to a logarithmic term. Because the inside of the log term should be dimensionless, there are no unique ways to remove it. However, the renormalized entanglement entropy is finite and does not depend on the cut-off. Thus, it has unique descriptions.  

We compute the renormalized entanglement entropy, corresponding to DOF at an energy scale $E\sim 1/l$. 
 According to \eqref{SUV2}, the $4d$ renormalized entanglement entropy becomes
\ba\label{REN23}
S_{ren}=L_4(l\partial_l)S=\dfrac{1}{2}(l\partial_l+1)(l\partial_l-1)S=\dfrac{1}{2}(l^2 S''+l S'-S),
\ea
where we have used the commutation relation $[\partial_l , l\partial_l ]=\partial_l$. Recall that the central charge of $d=4$ $\mathcal{N}=4$ SYM is $a={\pi^5 L^8 }/{\kappa_{10}^2}={N^2}/{4}$, where $\kappa_{10}^2=\pi^3 L^5 \kappa^2$. The renormalized entanglement entropy depends on the entangling surface and the trace anomaly~\cite{Liu:2012eea} as follows:
\ba \label{ano}
S_{ren}^{d=4}=2a_4 \int_{\partial A}d^2x \sqrt{h} E_2+c_4\int_{\partial A} d^2x \sqrt{h}I_2,
\ea
where $\partial A$ is the entangling surface (see also~\cite{Ryu:2006ef,Solodukhin:2008dh}).  In 4 dimensions, we have an A-type anomaly $a_4$ and a B-type anomaly $c_4$ on the entangling surface. $E_2$ is the Euler density and $I_2$ is a Weyl invariant. For the spherical entangling surface $\partial A =S^2$, $\int d^2x \sqrt{h} E_2=2$ and the Weyl invariant is zero. Thus, $S_{ren}^{d=4}=4a_4$. The renormalized entanglement entropy will satisfy the C-theorem in that case. Because the entangling surface is $S^1\times S^1$ for QFT on $R^{1,2}\times S^1$, however, the Euler number is zero. Only the B-type anomaly remains. The renormalized entanglement entropy can be non-monotonic since there is no universal C-theorem for B-type anomalies. 

\begin{figure}[htbp]
     \begin{center}
\includegraphics[height=4cm,clip]{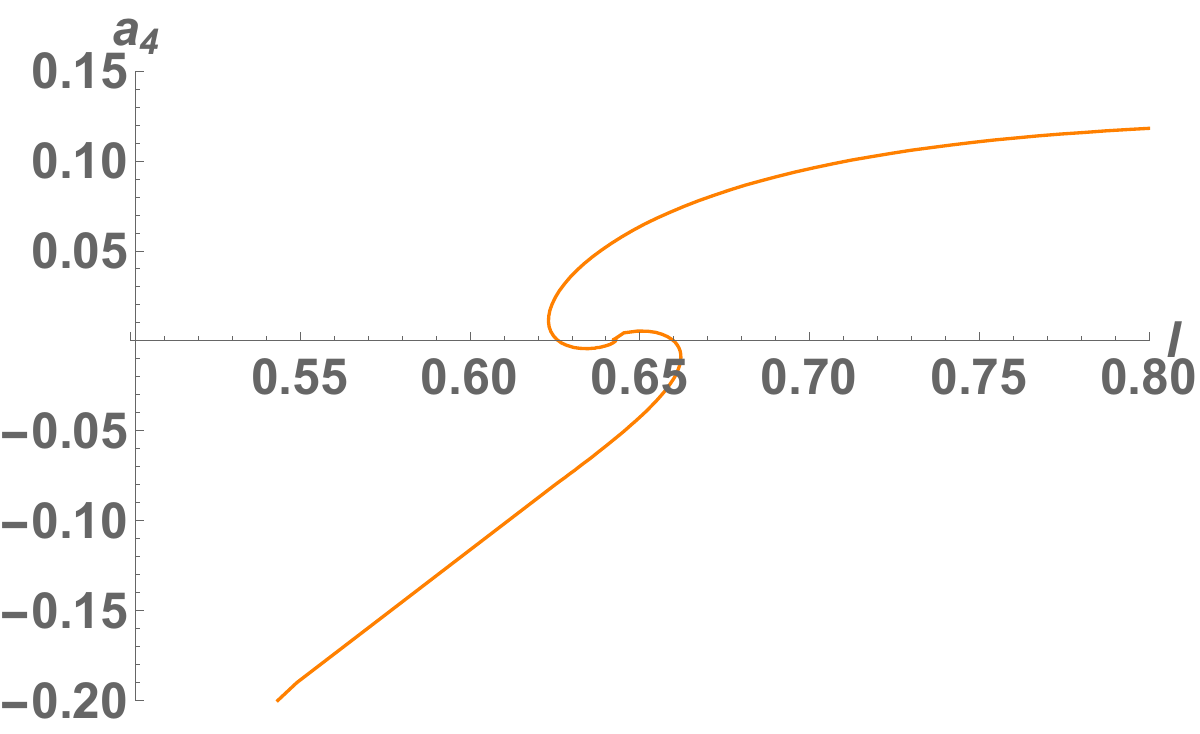}
\includegraphics[height=4cm,clip]{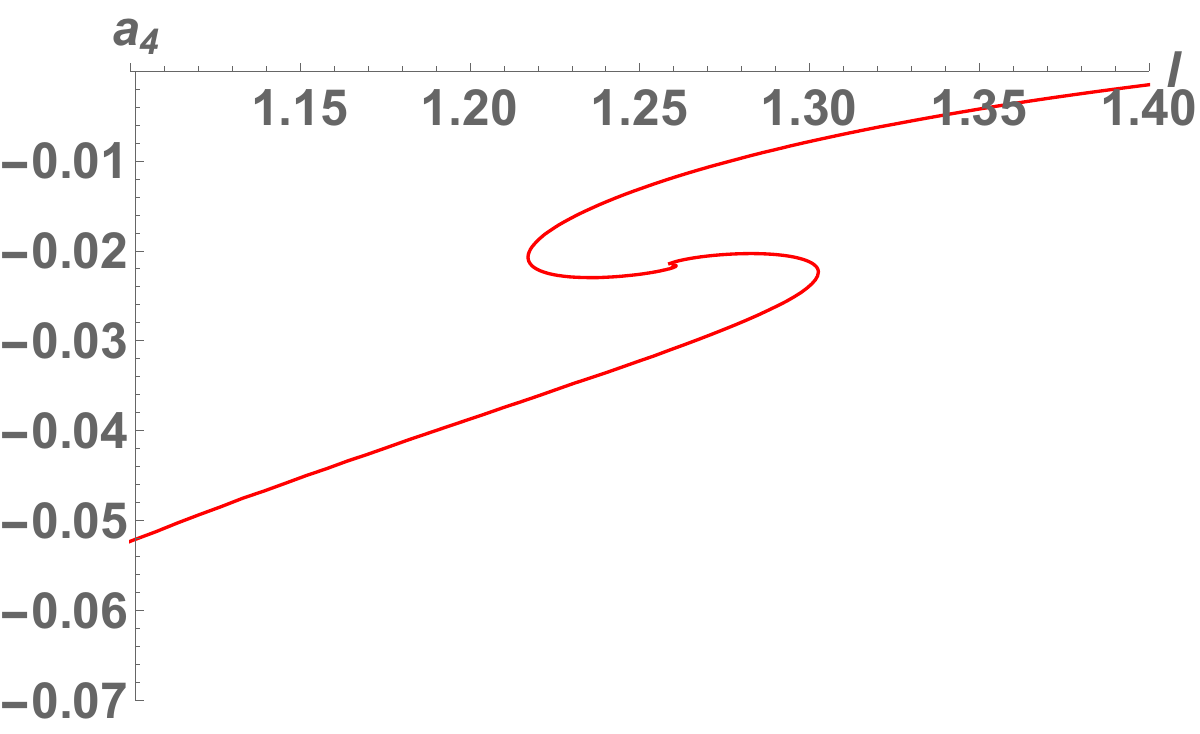}
\caption{ $a_4$ as a function of $l$. The disk shape dominates the behavior for small $l$, while the cylinder shape dominates the behavior for large $l$. Left: $a_{\phi}=\frac{i}{2}$. The critical length of the phase transition is $l_c=0.66$.  Right: $a_{\phi}=\frac{1}{\sqrt{2}}$. The critical length is $l_c=1.26$. }
    \label{fig:a4}
    \end{center}
\end{figure}

To compute the renormalized entanglement entropy, we need $a_4(l)$, $l$ appearing in \eqref{EXP} and $S$ (not confused with anomaly $a_4$ in (\ref{ano})). A numerical result of $a_4(l)$ is obtained in Fig.~\ref{fig:a4}. For pure imaginary $a_{\phi}$, it leads to results of geometric entropy~\cite{Allahbakhshi:2013wk}. The geometric entropy is related to entanglement entropy via the double Wick rotation~\cite{Bah:2008cj, Fujita:2008zv}. The disk shape is dominant for small $l$, and the cylinder shape is dominant for large $l$.  $a_4(l)$ becomes multi-valued near the phase transition at a critical length between disk and cylinder surfaces. Multi-valued behavior is also observed for other $a_{\phi}$.

\begin{figure}[htbp]
     \begin{center}
          \includegraphics[height=4cm,clip]{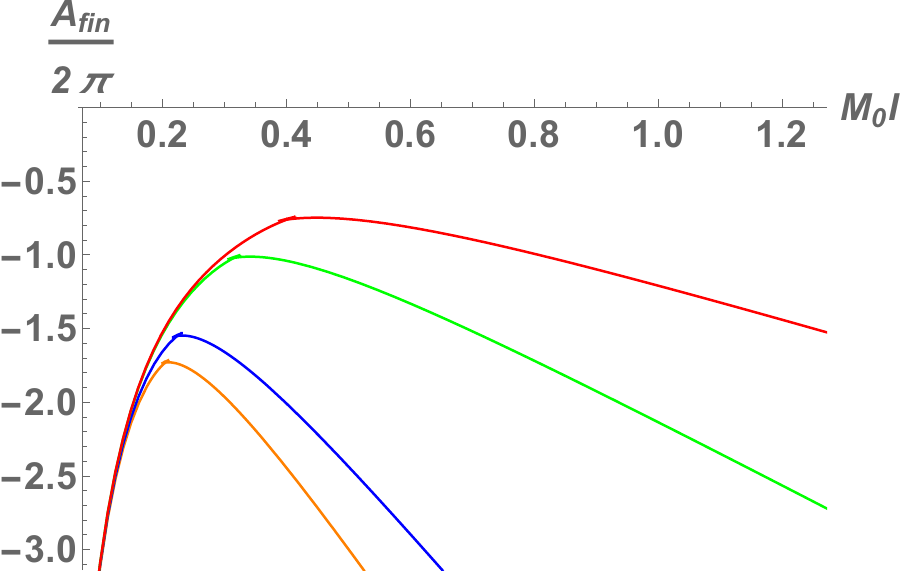}
          \includegraphics[height=4cm,clip]{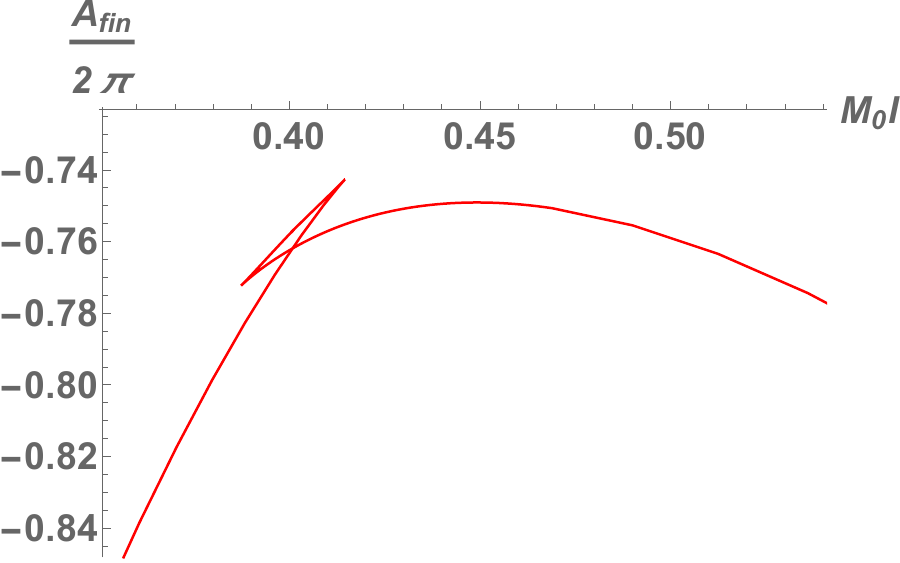}
           \caption{Left: the finite part of on-shell action $A_{fin}$ for $a_{\phi}=\frac{i}{2},\ 0,\ \frac{2}{3},\ \frac{1}{\sqrt{2}}$.
The figure shows that the entanglement entropy increases with the Wilson line $a_{\phi}$ increase. The quantum phase transition happens when $M_0l_c=0.21,\ 0.23,\ 0.32,\ 0.4$.  Right: closed-up figure of $A_{fin}$ for $a_\phi =\frac{1}{\sqrt{2}}$. }
    \label{fig:DA}
    \end{center}
\end{figure}

\begin{figure}[htbp]
     \begin{center}
          \includegraphics[height=5.2cm,clip]{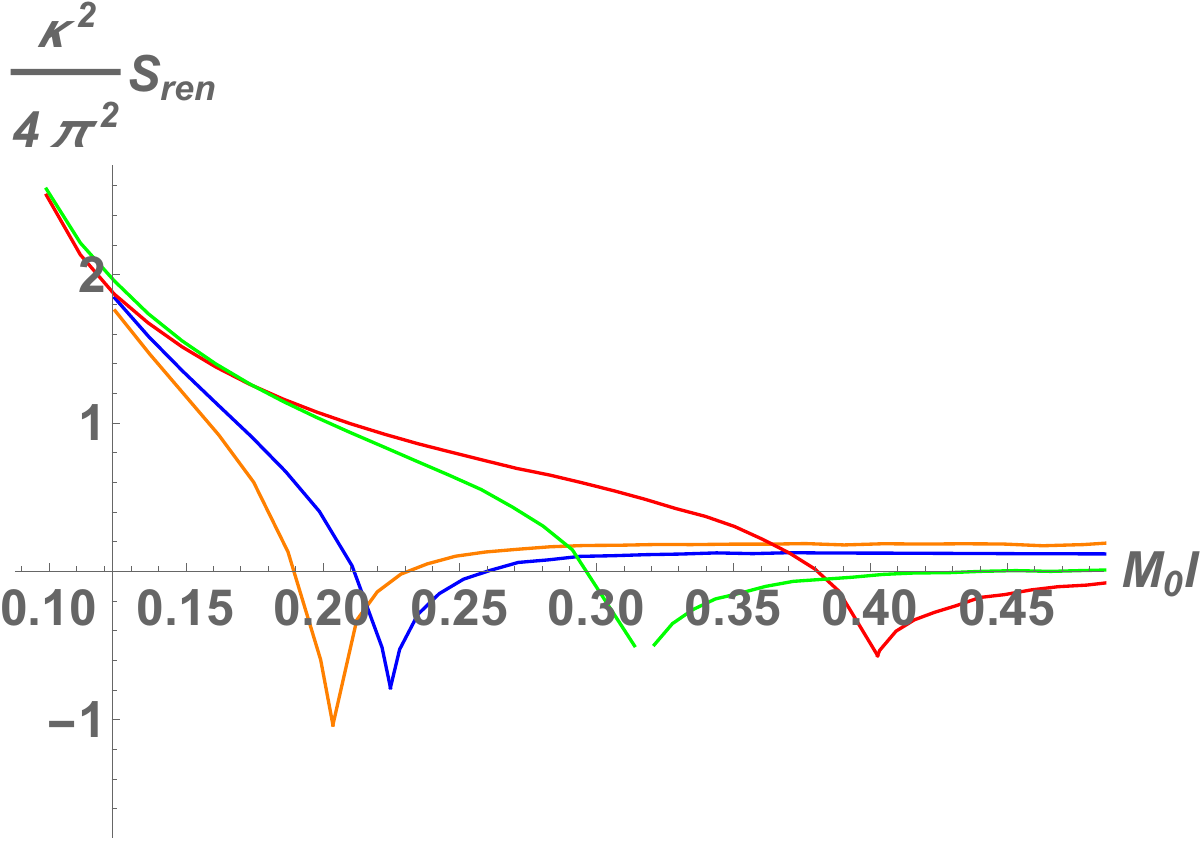}
           \caption{$S_{ren}$ for $d=4$. The renormalized entanglement entropy is plotted for several $a_{\phi}$. $a_{\phi}=\frac{i}{2},\ 0,\ \frac{2}{3},\ \frac{1}{\sqrt{2}}$ from the left to the right. The renormalized entanglement entropy non-monotonically behaves near critical lengths. The quantum phase transition happens when $M_0l_c=0.21,\ 0.23,\ 0.32,\ 0.4$. }
    \label{fig:Sren4}
    \end{center}
\end{figure}
\begin{figure}[htbp]
     \begin{center}
 \includegraphics[height=5.2cm,clip]{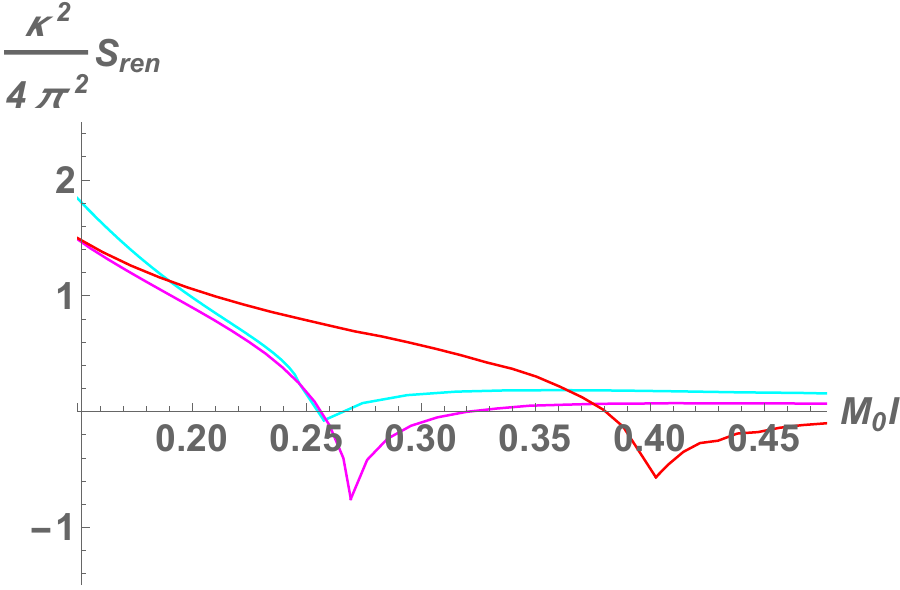}
           \caption{$S_{ren}$ for $d=4$. The renormalized entanglement entropy for several $M_0$: $M_0=1/\pi,\ 2/5,\ 3/5$ from the right to the left. The renormalized entanglement entropy non-monotonically behaves near critical lengths. The quantum phase transition occurs when $M_0l_c=0.24,\ 0.27,\ 0.4$. The figure shows that massive modes $Ml>1$ decouple others soon. The final states will be product states.}
    \label{fig:Sren4M0}
    \end{center}
\end{figure}

Substituting the boundary expansion \eqref{EXP} into the action \eqref{ACT3} and expanding at a small $z$, we obtain the following divergent part of $A=A_{fin}+A_{div}$:
\ba
\dfrac{A_{div}}{2\pi L_{\phi}}=\dfrac{l}{2\epsilon^2}+\dfrac{1}{8l}\log\Big(\dfrac{\epsilon }{l}\Big).
\ea
Thus, the divergent structure of entanglement entropy $A$ is
\ba
A=2\pi L_{\phi}\Big(\dfrac{l}{2\epsilon^2}+\dfrac{\log \epsilon}{8l}\Big)+S_{fin}(l),
\ea
where the log dependence is included in the finite part $S_{fin}$. $S_{fin}=A_{fin}-2\pi L_{\phi}\dfrac{1}{8l}\log l$.

To find minimal surfaces, one must compute the on-shell action of \eqref{ACT3}. Minimal surfaces between the disk and the cylinder dominate the phase in Fig. \ref{fig:DA}. The quantum phase transition occurs at a critical length $l_c$. Yellow and Blue curves show that the confinement occurs and decreases DOF~\cite{Witten:1998zw, Nishioka:2006gr, Klebanov:2007ws}.  Recall that $a_{\phi}$ increases Casimir energy of dual QFT, and then the entanglement entropy increases with the increase of $a_{\phi}$.

$S_{fin}$ is different from $S_{ren}$ because the cut-off dependence is removed at $S_{ren}$. Actually, $S_{fin}$ is related to  $S_{ren}$ via
\ba\label{SRE27}
S_{ren}=\dfrac{1}{2}(l^2 S''_{fin}+l S'_{fin}-S_{fin}).
\ea
Considering \eqref{dAl21}, $S_{fin}(l)$ satisfies the following relation:
\ba\label{SFI26}
S_{fin}'(l)=2\pi L_{\phi}\Big(-4la_4(l)-\dfrac{3}{32l^2}+\dfrac{1}{8l^2}\log (l)\Big).
\ea
Due to \eqref{SFI26}, $S''_{fin}$ or $A_{fin}''$ becomes
\be\bal\label{AFI29}
\dfrac{S_{fin}''}{2\pi L_{\phi}}=&\dfrac{5}{16l^3}-\dfrac{\log (l)}{4l^3}-4la_4'(l)-4a_4(l), \\
\dfrac{A_{fin}''}{2\pi L_{\phi}}=&-\dfrac{1}{16l^3}-4la_4'(l)-4a_4(l).
\eal\ee
Recall that \eqref{AFI29} is the finite part of the minimal surface $A$. The renormalized entanglement entropy $S_{ren}$ is finite. Substituting \eqref{SFI26} and \eqref{AFI29} into \eqref{SRE27}, $S_{ren}$ is rewritten as follows:
%\ba
%\dfrac{S_{ren}}{2\pi L_{\phi}}=-4 l^2 a_4(l)-2 l^3 a_4'(l)+\dfrac{7}{64l}-\dfrac{A_{fin}}{2}.
%\ea
\ba\label{SRR28}
&\dfrac{\kappa^2 S_{ren}}{4\pi^2 L_{\phi}}=-4 l^2 a_4(l)-2 l^3 a_4'(l)+\dfrac{7}{64l}-\dfrac{A_{fin}}{2} \nonumber \\
&=-4 l^2 a_4(l)-2 l^3 a_4'(l)+2\int^l l' a_4(l') dl'+\dfrac{1}{8l}+c_r,
\ea
The coefficient of $1/l$ comes from only the logarithmic term of $S$, which is brought from the Weyl anomaly. 
{The formula \eqref{SRR28} also }depends on a function of $a_4(l)$ unlike the holographic entanglement entropy of the spherical entangling surface in 4 dimensions~\cite{Ryu:2006ef}.  For the spherical entangling surface, here, HREE describes the A type anomaly in CFT: $S_{ren}^{d=4}=4a_4$. Recall that QFT dual to the AdS soliton with gauge potential breaks conformal invariance. The first three terms in \eqref{SRR28} represent terms breaking conformal invariance and the last term is to realize the small $l$ limit of HREE (CFT behavior)~\cite{Solodukhin:2008dh}. 

Fig. \ref{fig:Sren4} shows the renormalized entanglement entropy $S_{ren}$ for several $a_{\phi}$.
It becomes non-monotonic behavior, which is similar to a behavior of GPPZ flow~\cite{Liu:2012eea}.
Intuitively, the renormalized entanglement entropy is also a detector of the effective DOF of entangling states at the energy scale $El\sim 1$. For large energy $E/M_0\sim 1/(M_0 l)>1$, $S_{ren}$ decreases as a function of $lM_0$. Because the degrees of freedom with Wilson lines contribute to large $a_{\phi}$ and energy, however, $S_{ren}$ slowly decreases until the critical length (see green and red curves). For small  $E/M_0<1$, the renormalized entanglement entropy can not detect effective DOF and can almost become a constant. Even if the renormalized entanglement entropy increases after the quantum phase transition, it satisfies a kind of C-theorem: $S_{ren}(l\to 0)>S_{ren}(l\to \infty)$.

\subsection{$d=5$}
\begin{figure}[htbp]
     \begin{center}
 \includegraphics[height=7cm,clip]{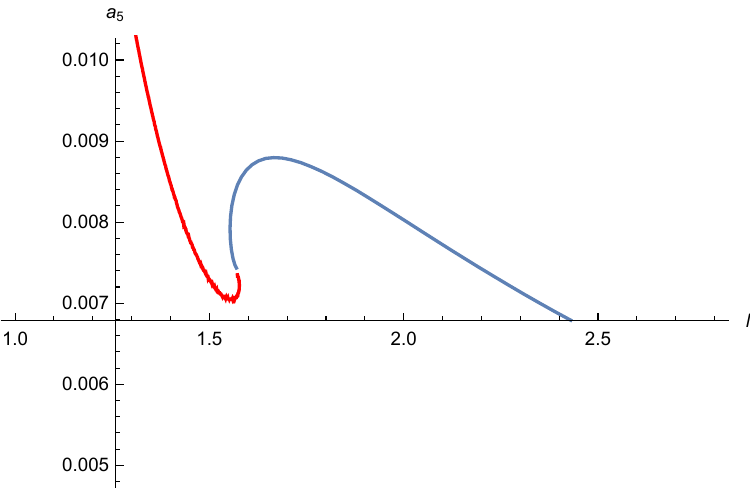}
\caption{$a_5$ as function of $l$ with $T=\frac{1}{\pi},a_\phi=\frac{1}{2}$ in the case of $d$=5. The red line corresponds to disk shaped RT surface and the blue line corresponds to a cylinder shape. The critical length of the phase transition is $l_c=1.57$.   }
    \label{aperon}
    \end{center}
\end{figure}

We proceed  with $d=5$ case, where the AdS boundary expansion for $d=5$ has the following form
\ba\label{RZ29}
R(z)=l-\dfrac{z^2}{3l}-\dfrac{5 z^4}{54 l^3} +z^5 a_5(l)\dots,
\ea
Similar to the case $a_4(l)$ discussed in the previous section, the parameter $a_5(l)$ is not determined from the AdS boundary expansion but determined from the IR boundary condition. Numerically, $a_5(l)$ is plotted in Fig.~\ref{aperon} for fixed values of $M_0$ and $a_{\phi}$.
Once we have the expansion \eqref{RZ29}, by making use of  \eqref{HJ7}, we can obtain the following derivative 
\ba\label{dAl30}
\frac{1}{\Omega_2L_{\phi}}\dfrac{dA}{dl}=\dfrac{2l}{3\epsilon^3}-5l^2 a_5(l),
\ea
where $\Omega_2=4\pi$. The first term is the cut-off dependent term. There are no log divergences. The finite part of $dA/dl$ is determined by the second term, which is similar to $d=3$ cases: a term corresponding to trace anomaly is not present in odd dimensions.  

\begin{figure}[htbp]
     \begin{center}
          \includegraphics[height=4cm,clip]{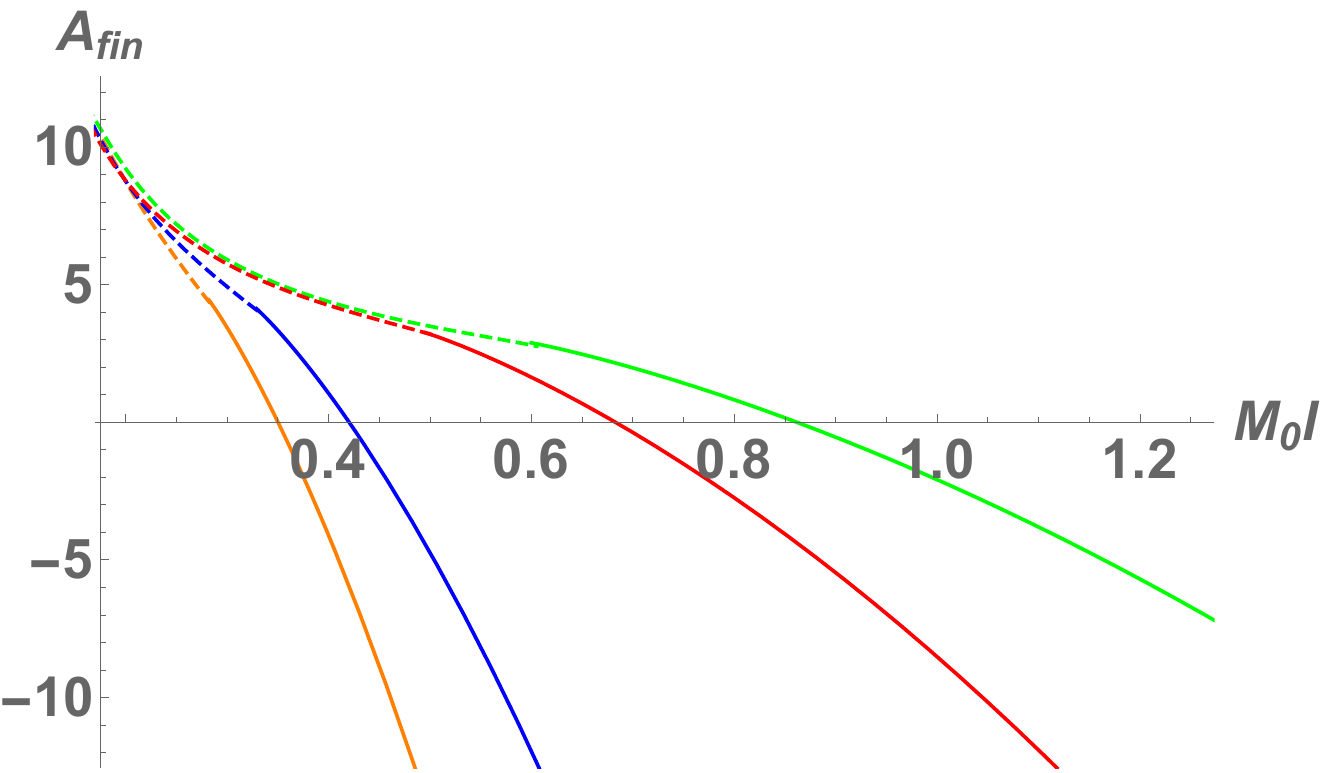}
          ~~~~\includegraphics[height=4cm,clip]{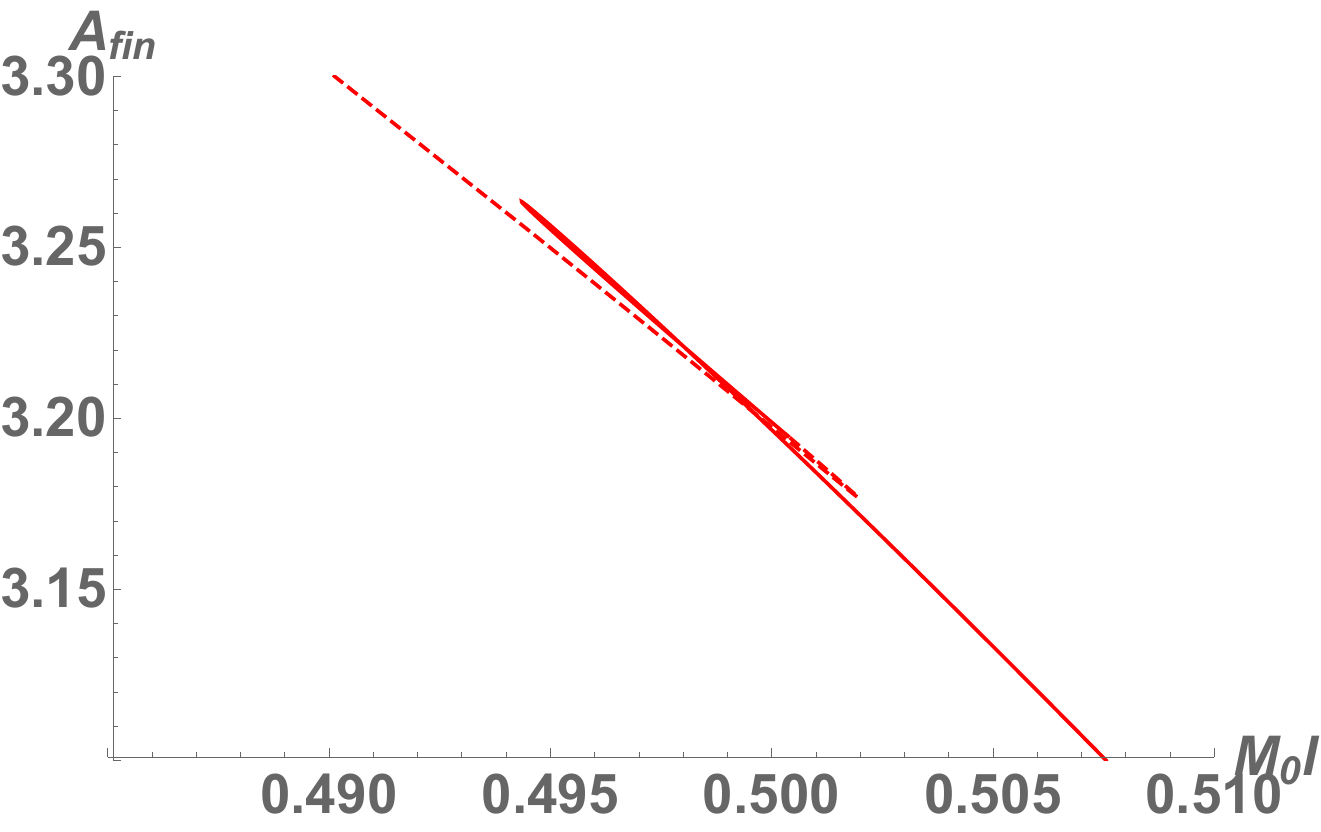}
           \caption{ Left: the finite part of $A$ for $a_{\phi}=i/2,\ 0,\ 1/2,\ 2/\sqrt{15}$ from left to right. Right: close-up version of $a_{\phi}=1/2$ curve. There is a phase transition at the critical length.}
    \label{fig:SEEd=5}
    \end{center}
\end{figure}
\begin{figure}[htbp]
     \begin{center}
          \includegraphics[height=4.5cm,clip]{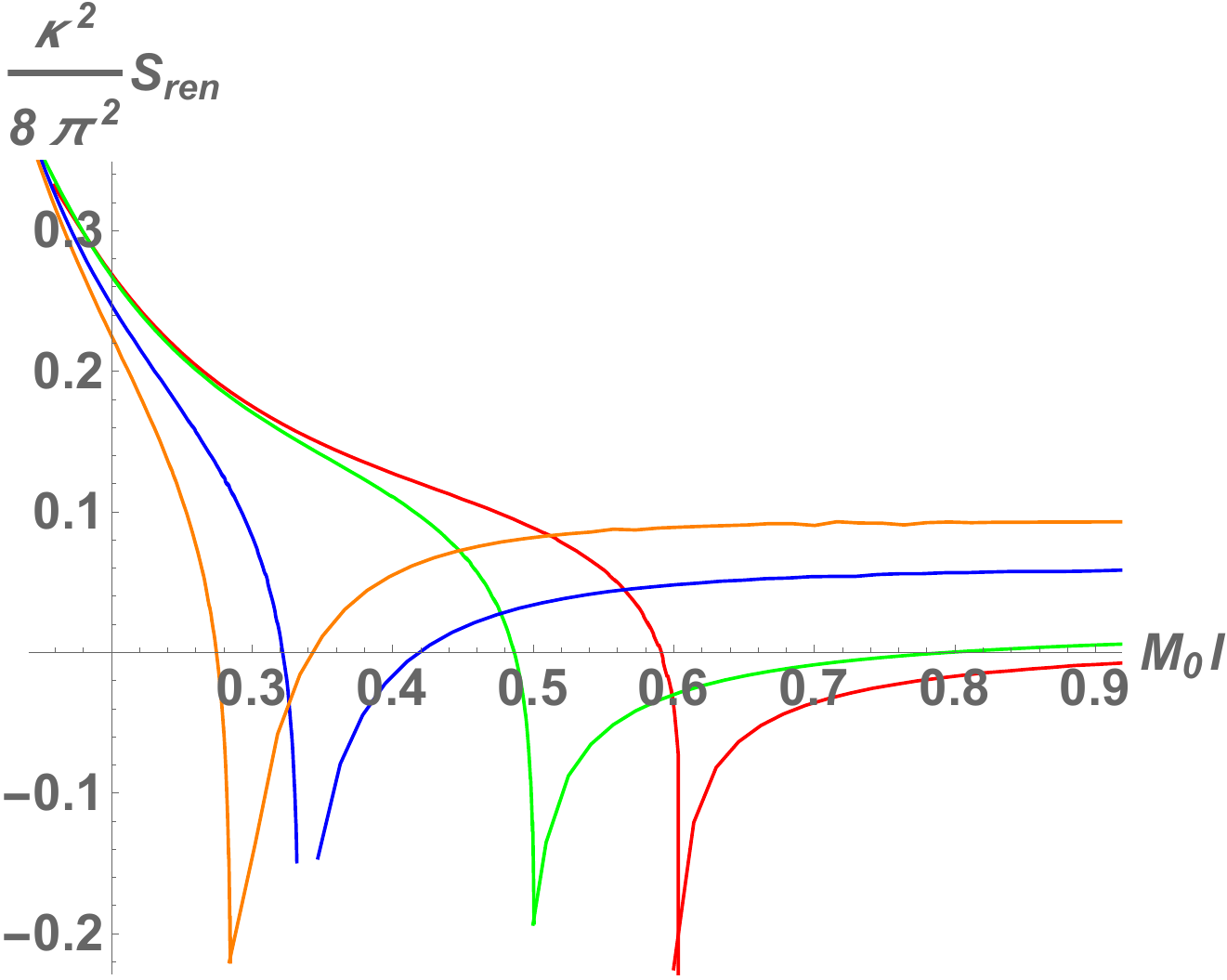}
          ~~~~\includegraphics[height=4.5cm,clip]{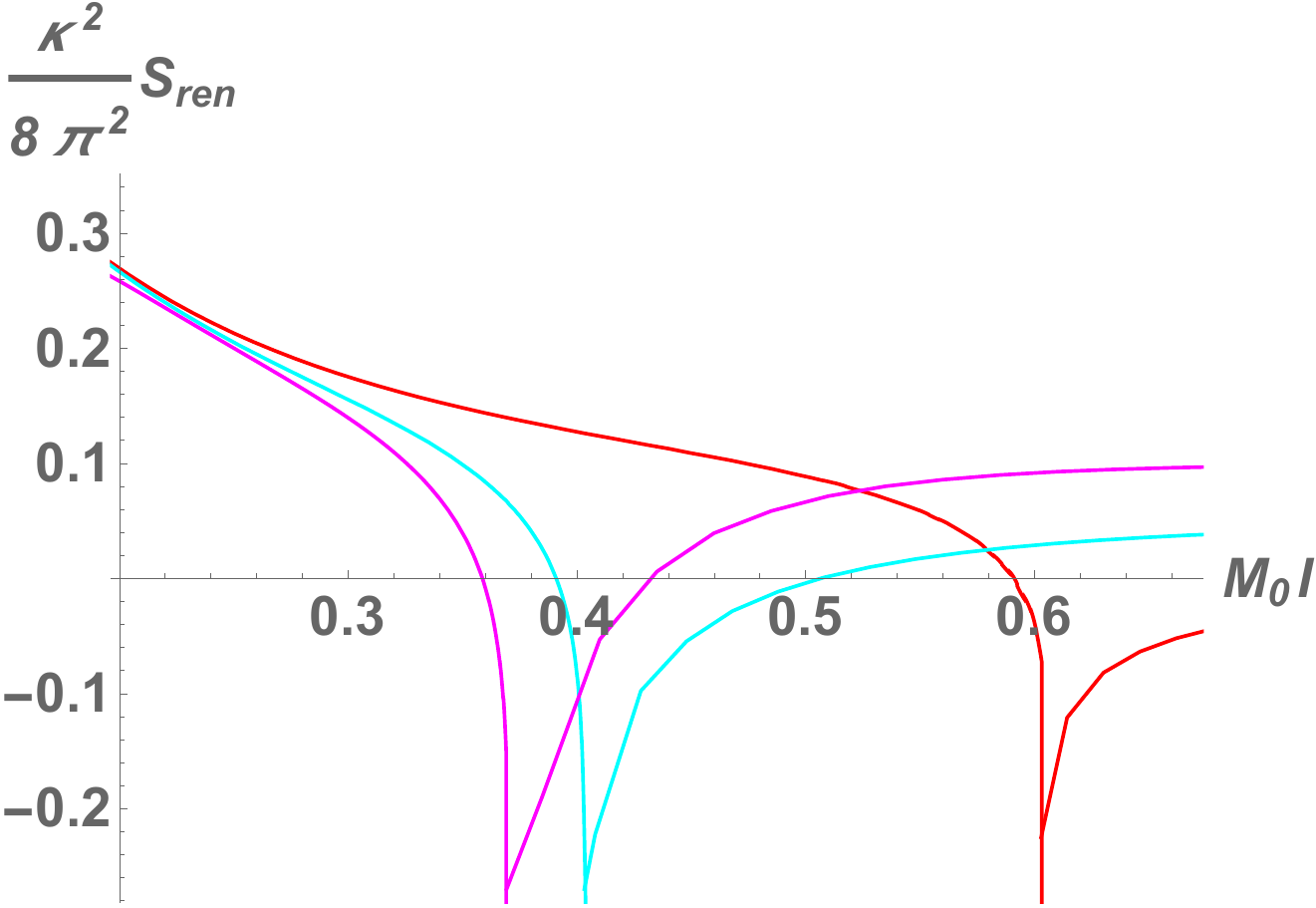}
           \caption{Left: The renormalized entanglement entropy for $a_{\phi}=i/2,\ 0,\ 1/2,\ 2/\sqrt{15}$. Renormalized entanglement entropy non-monotonically behaves. For large $a_{\phi}$, DOF slowly decreases. Right: The renormalized EE for $a_{\phi}=2/\sqrt{15}$. Mass is changed from $M_0=1/2,\ 2/5,\ 1/\pi$. Massive modes quickly decrease.}
    \label{fig:Srend=5}
    \end{center}
\end{figure}

%For small $l$, disk shaped RT surface dominates the behavior, while cylinder shape surface dominates for large $l$. The phase transition happens at a critical limit. 

Substituting the expansion \eqref{RZ29} into the action \eqref{ACT3} and expanding in terms of small $z$, we can obtain the divergent part of entanglement entropy which has the following divergent structure
\ba\label{Al31}
A(l)=4\pi L_{\phi}\Big(\dfrac{l^2}{3\epsilon^3}-\dfrac{4}{9\epsilon}\Big)+A_{fin}(l).
\ea
Since the second term in the parentheses does not depend on $l$, $O(\epsilon^{-1})$ part is absent in \eqref{dAl30}. The $l$ dependence of $A(l)$ differs from the CFT one.  We plotted the finite part of the entanglement entropy in Fig. \ref{fig:SEEd=5}, where it is shown that the disk-shaped RT surface dominates the behavior for small $l$ and cylinder shape dominates for large $l$. The finite part $A_{fin}$ has the quantum phase transition at the critical length. Making use of \eqref{dAl30} and \eqref{Al31}, the finite part satisfies the following relation
\ba\label{AFI25}
\dfrac{A_{fin}'(l)}{4\pi L_{\phi}}=-5l^2 a_5(l).
\ea
 
Next consider the 5-dimensional renormalized entanglement entropy defined in \eqref{SUV2}, which is  
\ba
S_{ren}=\dfrac{1}{3}l\partial_l (l\partial_l -2)S.
\ea 
Substituting \eqref{Al31}, the renormalized entanglement entropy can be rewritten as
\ba
S_{ren}(l)=\dfrac{l}{3}(-A_{fin}'(l)+l A_{fin}''(l))=-\dfrac{5}{3}\cdot 4\pi L_{\phi}\cdot l^3 (a_5(l)+l a_5'(l)).
\ea
The renormalized entanglement entropy is plotted in Fig. \ref{fig:Srend=5} with {different $a_\phi$}.  The renormalized entanglement entropy counts the DOF of the entangling states at an energy scale $E\sim 1/l$. It has non-monotonic behaviors. The quantum phase transition happens at critical limits. And $S_{ren}$ approaches to zero at large distances. Fig. \ref{fig:Srend=5} (right) shows that massive DOF decouples other modes at low energy soon.  This behavior is easily seen when $a_{\phi}=0$, where scaling symmetry $M_0\to \lambda M_0,\ l\to \lambda^{-1}l$ arises. This is the symmetry of the action and EOM. Scaling symmetry simultaneously rotates $a_{\phi}$ when it is nonzero. The scaling symmetry implies that the critical length is inversely proportional to $M_0$: $l_c =c_5/M_0$. The quantum phase transition quickly occurs for large masses, while it slowly occurs for small masses. After the phase transition, the renormalized entanglement entropy gradually becomes {constant}.

\section{$d=3$ (a striped shape)}\label{a striped shape d=3}
In this section, we analyze the holographic entanglement entropy for $d=3$. The configuration for $d=3$ is equal to a striped boundary shape. We start with $d$ dimensional striped shapes, and $d=3$ is a special case. To consider a striped shape, we replace $R^2d\Omega_{d-3}$ with $\sum dx_{\perp}^2$ in  $d+1$ dimensional $AdS$ soliton with a gauge field~\eqref{METd1} as follows: 
\ba\label{METd2}
&ds_{d+1}^2=\dfrac{L^2}{z^2}\Big(\dfrac{dz^2}{f_d(z)}+f_d(z)d\phi^2-dt^2+dR^2+\sum dx_{\perp}^2\Big), \nonumber 
\ea
where $R$ is along ($-\infty, \ \infty$) unlike the radial direction of polar coordinate systems.
\begin{figure}[htbp]
     \begin{center}
          \includegraphics[height=5cm,clip]{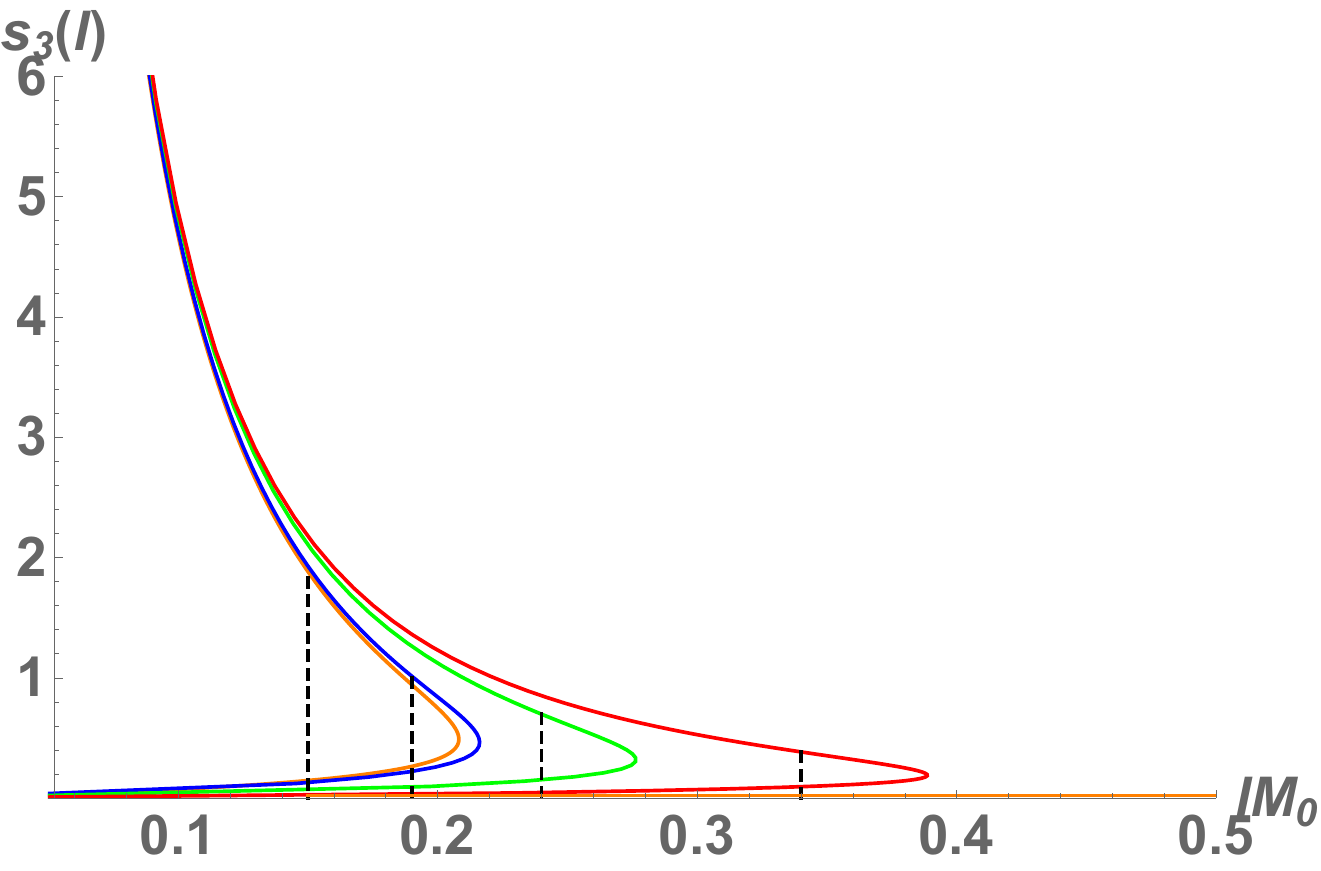}
          \includegraphics[height=5cm,clip]{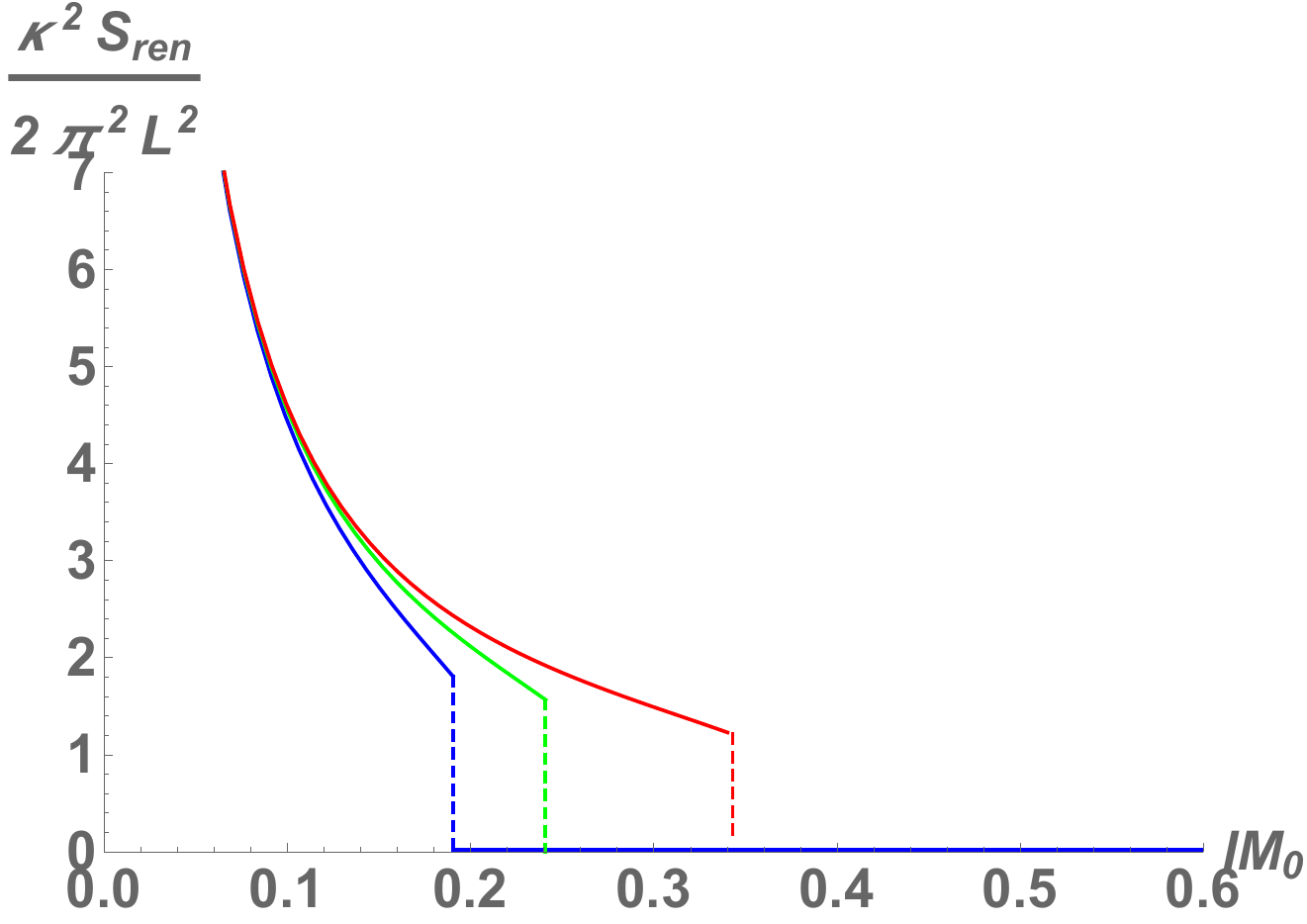}
           \caption{Left: $s_3(l)=dA_s/(3L_{\phi}dl)$ as a function of $lM_0$. The curve is for $a_{\phi}=i/2,\ 0,\ 1,\ 2/\sqrt{3}$ from the left to the right.  The phase transition happens at critical lengths $l_c=0.15,\ 0.19,\ 0.24,\ 0.34$ in units of $1/M_0$ from the left to the right, respectively. $s_3(l)$ has two values. The physical curve (an upper one) is consistent with the strong subadditivity. Right: $S_{ren}$ for $d=3$. The curve is for $a_{\phi}=0,\ 1,\ 2/\sqrt{3}$ from the left to the right.  $S_{ren}$ monotonically decreases as a function of $lM_0$. $S_{ren}$ becomes 0 at a large distance. It implies that theory is a product state there~\cite{Latorre:2004pk}.}
    \label{fig:Sren}
    \end{center}
\end{figure}
\begin{figure}
     \begin{center}
          \includegraphics[height=5.4cm,clip]{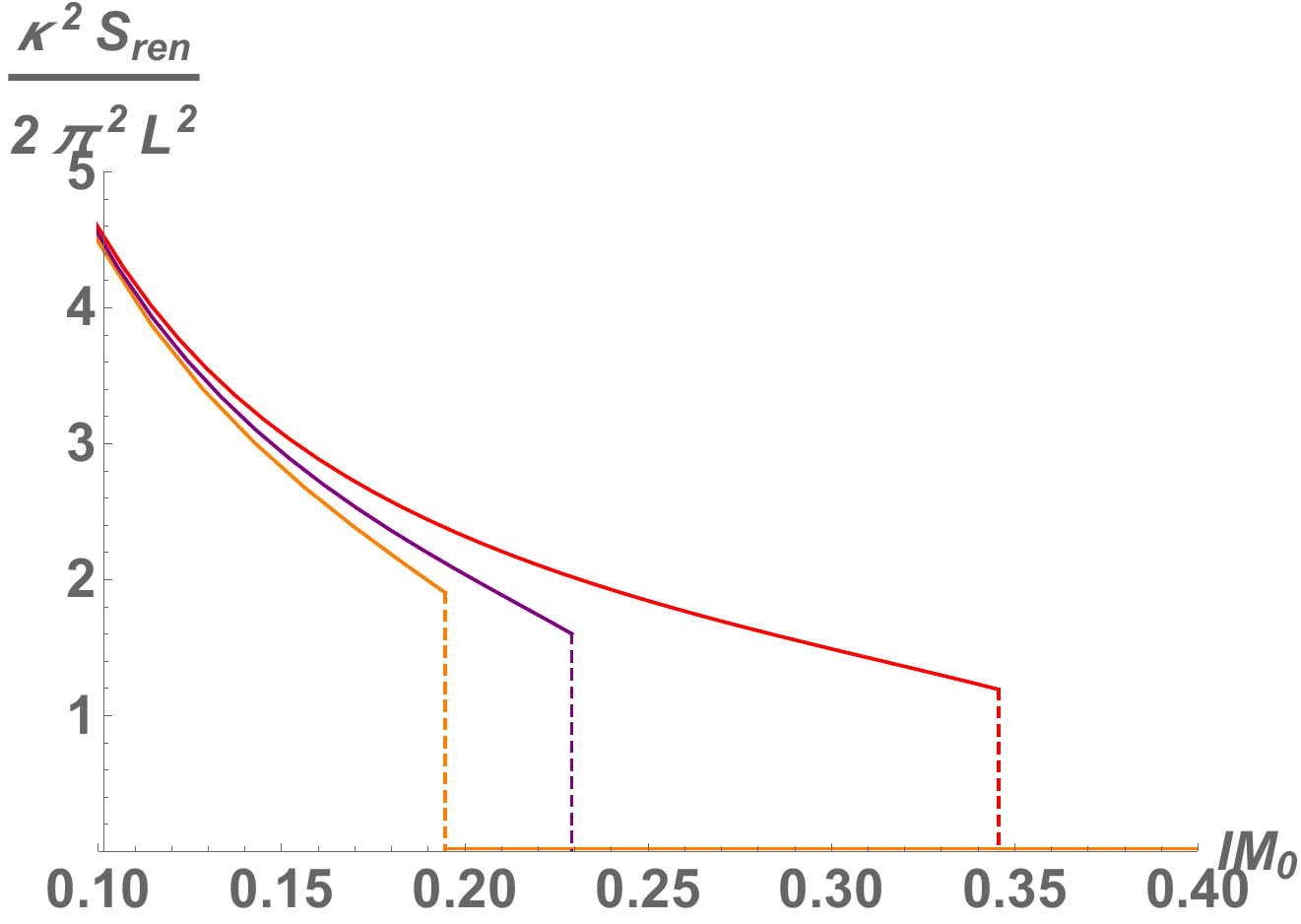}
           \caption{$S_{ren}$ for $d=3$. The curve is for $a_{\phi}=2/\sqrt{3}$. $M_0=3/5,\ 2/5,\ 1/\pi$ from the left to the right. $S_{ren}$ monotonically decreases as a function ozimf $lM_0$. The phase transition happens at critical lengths $l_cM_0=0.19,\ 0.23,\ 0.34$ from the left to the right, respectively. It implies that massive modes $M>1/l$ quickly decouple others and remain product states. }
    \label{fig:Mch}
    \end{center}
\end{figure}

We consider a strip with a length of $l$ along the $R$ direction ($-l/2 \le R \le l/2$) and choose $R=R(z)$ as an embedding scalar. 
The surface action becomes 
\ba\label{ACTst}
A_s=\int d^{d-1}x\mathcal{L}=2V_{d-3}L_{\phi}{L^{d-1}}\int dz \dfrac{1}{z^{d-1}}\sqrt{1+f \dot{R}^2}, 
\ea
where  $V_{d-3}$ was used to replace the volume of $d-3$ dimensional space spanned by $x_{\perp}$. Factor 2 comes from two contributions of the minimal surface. Note that the lagrangian density of \eqref{ACTst} does not depend explicitly on $R$. This simplifies the analysis for a rectangular region more than for a circular region. The momentum doesn't explicitly depend on $R$ for a striped boundary shape. Solving the {condition $\Pi=$const and} imposing the IR boundary condition $dz/dR|_{z=z_t}=0$, we have
\ba\label{EOM11}
\dot{R}=\dfrac{1}{\sqrt{f_d(z) \Big(\dfrac{f_d(z) z_t^{2(d-1)}}{f_d(z_t)z^{2(d-1)}}-1\Big)}}.
\ea 
This formula demonstrates that $z=z_t$ is the turning point  $dz/dR|_{z=z_t}=0$. The equation \eqref{EOM11} gives the profile of the minimal surface satisfying $R(\epsilon)=-l/2$ and $R(z_t)=0$. 

Using the $AdS$ boundary expansion $R'(z)=d s_d(l) z^{d-1}+\dots$, $R(z)$ is expanded as follows
\ba
R(z)=-\dfrac{l}{2}+s_d(l) z^d\dots ,
\ea
where 
\ba
s_d(l)=\dfrac{\sqrt{f_d(z_t)}}{dz_t^{d-1}}.
\ea

The Hamilton-Jacobi equation for a striped shape has the same form as \eqref{HJ7}
\ba\label{HJs}
\dfrac{dA_s}{dl}=-\Pi (\epsilon)\dfrac{dR (\epsilon)}{dl},
\ea
where 
\ba
\Pi =\dfrac{\partial \mathcal{L}}{\partial \dot{R}}=2V_{d-3}L^{d-1}L_{\phi}\dfrac{f\dot{R}}{z^{d-1}\sqrt{1+f\dot{R}^2}}
\ea
and $H(z_t)=0$ due to the IR boundary condition. When \eqref{EOM11} is substituted into \eqref{HJ7}, the $l$ derivative of the surface becomes
\ba\label{ENTd1}
\dfrac{dA_s}{V_{d-3}L_{\phi}L^{d-1}dl}=d s_d(l)=\dfrac{\sqrt{f_d(z_t)}}{z_t^{d-1}}.
\ea
Note that the minimal surface must satisfy strong subadditivity. It requires that the minimal surface must be concave~\cite{Hirata:2006jx} as follows:
\ba
A_s''=d V_{d-3}L_{\phi}L^{d-1}s_d'(l)=V_{d-3}L_{\phi}L^{d-1}\dfrac{\partial z_t}{\partial l}\partial_{z_t}(ds_d(l)) \le 0.
\ea
Actually, one can show the following inequality:
\ba
\partial_{z_t}(ds_d(l))\equiv \dfrac{h(z_t)}{2\sqrt{f_d(z_t)}}=\dfrac{(-2(d-1)z_t^{-d}+(d-2)(1-a_{\phi}^2z_0^2)z_0^{-d})}{2\sqrt{f_d(z_t)}}<0.
\ea
because $h(z)$ satisfies the following condition:~\footnote{ When $a_{\phi}$ is pure imaginary $a_{\phi}=ia_{\phi}'$, the second condition is replaced with
\ba
h(z_0)=-\dfrac{4\pi (\sqrt{4\pi^2 +d(d-2)(a_{\phi}'/M_0)^2}-2\pi)}{(a_{\phi}'/M_{0})^2(d-2)}<0.
\ea}
\ba
h(0)=-\infty,\quad h(z_0)=(-d-(d-2)(a_{\phi}z_0)^2)z_0^{-d}<0,\quad h'(z)=2d(d-1)z^{-d-1}>0.
\ea
According to~\cite{Fujita:2020qvp}, in addition, $\partial z_t/\partial l$ has two values. It is positive for a minimal surface and negative for an unphysical curve.

Next, we consider the case in which the minimal surface is disconnected. We focus on $d=3$. 
From
\ba\label{eom_disconnected}
\frac{\mathrm{d} }{\mathrm{d} z}\frac{\partial \mathcal{L} }{\partial \dot{R} } =\frac{\partial \mathcal{L} }{\partial R} =0
\ea
where
\ba
\mathcal{L}\equiv \frac{1}{z^{2} } \sqrt{1+f\dot{R}^2 }, 
\ea
it is easy to see that one of the solutions is $R=\text{const}$, which is for the disconnected minimal surface. Then the area of the surface is
\ba
A_{s}=2L_{\phi } \int_{\epsilon }^{z_{t} }\mathrm{d}z \frac{1}{z^{2} } 
       =2L_{\phi }\left ( \frac{1}{\epsilon } -\frac{1}{z_{t} } \right ), 
\ea
and the corresponding entanglement entropy
\ba
S_{EE}=\frac{ 4\pi L_{\phi }}{\kappa } \left ( \frac{1}{\epsilon } -\frac{1}{z_{t} } \right ),
\ea
where $z_t=z_+$. The finite part of $S_{EE}$ is $-4\pi L_{\phi}/(\kappa z_+)$. For disconnected surfaces, $s_3(l)=0$.  

One can compute the renormalized entanglement entropy from the entanglement entropy $S_{EE}$.
We apply the renormalized entanglement entropy formula defined in \eqref{SUV2}: $S_{ren}=l\partial_l S$  substituting $d=3$. Because dual 3-dimensional QFT theory is defined on $R^{1,1}\times S^1$, the renormalized entanglement entropy will give the result of $2d$ QFT on $R^{1,1}$ in the low energy limit.

 The renormalized entanglement entropy can be written in terms of $l$ and $s_3(l)$ as follows: 
\ba
\dfrac{\kappa^2}{2\pi L_{\phi}L^2}S_{ren}=\dfrac{ldA_s}{L_{\phi}L^2dl}=3 ls_3(l)=\dfrac{l\sqrt{f(z_t)}}{z_t^2},
\ea
where $S=\frac{2\pi}{\kappa^2}A_s$. Note that the central charge of $3d$ dual CFT is $32L^2/(\pi G_4) =2^6\sqrt{2}N^{3/2}\sqrt{k}/(3\pi)$, where $k$ is the Chern-Simons level. 
 For the disconnected surface, we see that the entanglement entropy does not depend on the size $l$ of the entangling surface at the boundary, which means
\ba
S_{ren}=l\partial_{l}S_{EE}=0.
\ea
 We plotted $dA_s/(3L_{\phi}dl)=s_3(l)$ in Fig. \ref{fig:Sren} (left). Because the concavity of the entanglement entropy \cite{Lieb:1973cp}\cite{Lieb:1973zz} means $A_s''\le 0$, $A_s'$ monotonically decreases as a function of $l$. We plotted the renormalized entanglement entropy in Fig.~\ref{fig:Sren} (right). The renormalized entanglement entropy detects the DOF of the entangling states at the energy scale $El\sim 1$ again. Because massive degrees of freedom decouple in the low energy limit, $S_{ren}$ decreases as a function of $l$ (with energy $E\sim 1/l$).~\footnote{Another entropic c function non-monotonically behaves in~\cite{Fujita:2020qvp} and it does not satisfy a c-theorem. The non-monotonic behavior occurs due to the competition between a power of $l$ and $dA_s/dl$.} This monotonic behavior ($S_{ren}'(l)\le 0$) will also be the consequence of Lorentz symmetry and strong subadditivity~\cite{Hirata:2006jx}. The quantum phase transition happens at critical lengths $l_cM_0=0.19,\ 0.24,\ 0.34$ for $a_{\phi}=0,\ 1,\ 2/\sqrt{3}$, respectively. $S_{ren}$ becomes 0 at large distances. Fig.~\ref{fig:Mch} changes $M_0$ after fixing $a_{\phi}$. It shows that massive modes $M_0>1/l$ quickly decouple others at low energy, remaining product states. When $a_{\phi}=0$, one can recover scaling symmetry $M_0\to \lambda M_0$, $l\to \lambda^{-1}l$, in addition to bulk parameters. This is the symmetry of the action and EOM. Scaling symmetry simultaneously changes $a_{\phi}$ if it is nonzero. Thus, the critical length is the same as those for different $M_0$ and is given by $M_0=c/l_c$, where $c$ is a constant. The phase transition happens soon for large masses and slowly for small masses.

Our results agree with the fact that the renormalized entanglement entropy is equal to the formula of the entropic $c$-function $l\partial_l S_{EE}$ on $R^{1,1}$, which becomes observable in renormalizable theory. The entropic $c$-function on $R^{1,1}$ counts DOF and depends on the radius $l (\sim 1/E)$ including the information in $S_{EE}(l_1)-S_{EE}(l_2)$ ($l_1>l_2$). It is known that the $c $ function for a massive scalar field exponentially decreases for large $r$~\cite{Casini:2004bw}. Finally, there remains no degree of freedom (zero entropic $c$ function) at a large distance. Recall that the trace anomaly exists in 2 dimensions. The monotonic quantity is the Euler term in the trace anomaly there.

\subsection{Small subregions}
By employing the our previous results, where $z_t$ can be expressed in terms of $l$ for small $l$   \cite{Fujita:2020qvp} 
\be\bal 
z_t=&\dfrac{l}{2 e_{-1}-2 k_{-1}}+\dfrac{l^4 \bar{a}_{\phi } \left(-4 e_{-1}+4 k_{-1}+\pi \right)}{256 z_+^3 \left(e_{-1}-k_{-1}\right){}^5}+\dfrac{l^5 \left(1-\bar{a}_{\phi }\right)}{160 z_+^4 \left(e_{-1}-k_{-1}\right){}^5}  \\
&+\dfrac{l^7 \bar{a}_{\phi }^2 \left(-302 e_{-1} k_{-1}+126 e_{-1}^2-84 \pi  e_{-1}+176 k_{-1}^2+84 \pi  k_{-1}+21 \pi ^2\right)}{172032 z_+^6 \left(e_{-1}-k_{-1}\right){}^9}+\dots,
\eal\ee
we obtain 
\be\bal\label{DAL16}
\dfrac{dA_s}{L_{\phi}{L^2}dl}=&\dfrac{4\left( e_{-1}- k_{-1}\right){}^2}{l^2}-\dfrac{l \pi  \bar{a}_{\phi }}{16 \left(z_+^3 \left(e_{-1}-k_{-1}\right){}^2\right)}+\dfrac{9 l^2 \left(\bar{a}_{\phi }-1\right)}{40 z_+^4 \left(e_{-1}-k_{-1}\right){}^2}   \\
&+\dfrac{5 l^4 \bar{a}_{\phi }^2 \left(80 e_{-1} k_{-1}-80 k_{-1}^2-21 \pi ^2\right)}{86016 z_+^6 \left(e_{-1}-k_{-1}\right){}^6}+\dots   \\
=&\dfrac{8 \pi ^3}{l^2 \Gamma \left(\frac{1}{4}\right)^4}-\dfrac{8 l \bar{a}_{\phi} \Gamma \left(\frac{5}{4}\right)^4}{\pi ^2 z_+^3}+\dfrac{9 (\bar{a}_{\phi}-1) l^2 \Gamma \left(\frac{1}{4}\right)^4}{80 \pi ^3 z_+^4}+\dfrac{5 (20-21 \pi ) \bar{a}_{\phi}^2 l^4 \Gamma \left(\frac{1}{4}\right)^{12}}{688128 \pi ^8 z_+^6}+\dots ,
\eal\ee
where we have defined $\bar{a}_\phi =1-(z_+a_{\phi})^2$. For the notation $e_{-1},k_{-1}$ please refer to \cite{Fujita:2020qvp}  for more details. 
The expansion \eqref{DAL16} agrees with the holographic entanglement entropy in \cite{Fujita:2020qvp}, and the formula \eqref{ENTd1} is also valid for other dimensions realizing the holographic entanglement entropy. The result is compared with our previous paper~\cite{Fujita:2020qvp} in the appendix \ref{appB}.

\section{Summary and discussion}
We computed holographic EE and the renormalized EE in the AdS soliton with the gauge potential for several dimensions. The disk shape of the minimal surface was dominant for small $l$, and the cylinder shape was dominant for large $l$ as similar to~\cite{Ishihara:2012jg}. The quantum phase transition occurs at a critical size of the subregion. 

The renormalized EE, a universal part of EE independent of the cutoff, was computed by operating differentiation on EE~\cite{Liu:2012eea}. By containing modes with KK mass and  considering the low energy limit, we continuously derived from odd dimensional renormalized EE to the formula of $d-1$ dimensional renormalized EE.~\footnote{Note that the topology of the subregion is not one ball $B^{d-1}$ but the ball and a circle $B^{d-2}\times S^1$ with the periodicity $L_{\phi}$.  Interpreting this $S^1$ as one perpendicular direction to $B^{d-2}$ with the endpoint identified will be convenient.} Actually, the $\phi$ circle shrinks to zero at the tip of the AdS soliton ($z=z_+$), which is probed for large $l$. The logarithmic term is absent since we don't have a Weyl anomaly in odd dimensions. This is a sort of topology change in the entanglement entropy. In any dimension, massive modes $M_0l>1$ decouple others as a decrease of energy as shown in Fig.~\ref{fig:Mch} and then product states are retained. For high energy limit  ($l\ll L_{\phi}$), the renormalized EE recovers behaviors of the original dimensions because the renormalized EE measures the degrees of freedom in a state with energy $E\sim 1/l$. Because the degrees of freedom with Wilson lines contribute to large $a_{\phi}$ and high energy, the renormalized EE slowly changes until the critical length (see Fig. \ref{fig:Sren}). The paper~\cite{Deddo:2022wxj} also tracked the entanglement entropy across dimensions and found transitions. 

In section \ref{a striped shape d=3}, we analyzed striped surfaces for $d=3$.  Our results demonstrated that the renormalized EE is positive (non-negative) and satisfies the C theorem. After dimensional reduction, the renormalized entanglement entropy of 2d QFT with Kaluza-Klein modes will also be consistent with the C theorem of the 2-dimensional entropic $c-$function. 
We showed that when $d=3$, the renormalized EE for the entangling boundary with a striped shape decreases monotonically and jumps to zero when the size of the entangling surface becomes large, which means that it probes a first-order phase transition and the corresponding boundary field theory runs into a product state at the low energy scale.  Unlike the generalized entropic $c-$function studied in \cite{Fujita:2020qvp}, the renormalized EE here behaves monotonically even for the large value of gauge potential  $a_{\phi}$, and since $a_{\phi}$ can increase the degrees of freedom and the increase of $a_{\phi}$ leads to the increase of the renormalized EE, which implies that the renormalized EE is counting the degrees of freedom (DOF) of the boundary field theory. 

 We analyzed the striped entangling surface in the previous analysis and computed an entropic $c-$function. This entropic $c-$function is always positive and non-monotonically behaves. The phase transition happens at a critical length. The non-monotonic behavior is caused by the effective DOF of Wilson lines along the $S^1$ direction. That is, effective DOF increases the renormalized entanglement entropy. It implies that this Wilson line will decrease the mass of particles such as glueballs~\cite{Csaki:1998qr} because particles of small mass contribute to the entropic $c-$function at large $l$. On the other hand, in higher dimensions, HREE can become negative near the phase transition point, different from an entropic $c-$function. This will be considered as an artifact in the large $N$ and strongly coupled limit. The two are similar in the quantum phase transition's presence and the Wilson lines' effect. The gauge potential will decrease the mass of particles such as glueballs. Thus, more DOF will contribute to HREE and let it increase at a large distance. 

\section*{Acknowledgments}
 We would like to thank X. Chen and P. Zhang for their helpful discussion. S.H. would appreciate the financial support from the Fundamental Research Funds for the Central Universities and Max Planck Partner Group and the Natural Science Foundation of China (NSFC) Grants No.~12075101 and No.~12235016. This work is also supported by the National Natural Science Foundation of China (No.12105113)

\appendix
\section{Hamilton-Jacobi equations}\label{HJmethod}
In the appendix, we give a brief review of the Hamilton-Jacobi method used in analyzing minimal surfaces in \eqref{HJ7}.
We introduce the following action 
\ba
S=\int^{t_2}_{t_1}\mathcal{L}(q,\dot{q},t)dt.
\ea
We assume that the field $q$ can change at the boundary times $t_1$ and $t_2$. Moreover, we allow changes of times $t_1(t_2)$ into $t_1'(t_2')$, respectively. The variation of the action becomes
\be\bal
\delta S =&\int^{t_2'}_{t_1'}dt\mathcal{L}(q',\dot{q}',t)dt-\int^{t_2}_{t_1}dt\mathcal{L}(q,\dot{q},t)dt \nonumber \\
=&\int^{t_2'}_{t_2}\mathcal{L}(q',\dot{q}',t)dt+\int^{t_2}_{t_1}(\mathcal{L}(q',\dot{q}',t)-\mathcal{L}(q,\dot{q},t))dt  +\int^{t_1}_{t_1'}\mathcal{L}(q',\dot{q}',t)dt \nonumber \\
=&\mathcal{L}(q',\dot{q}',t_2)\delta t_2-\mathcal{L}(q',\dot{q}',t_1)\delta t_1+\dfrac{\partial \mathcal{L}}{\partial \dot{q}}\delta q\Big|_{t_2}-\dfrac{\partial \mathcal{L}}{\partial \dot{q}}\delta q\Big|_{t_1},
\eal\ee
where the EOM is used in the second line. We did not assume $\delta q =0$ at the time boundary $t_1$ and $t_2$.
We use the following transformation
\ba
\dfrac{\partial \mathcal{L}}{\partial \dot{q}}\delta q\Big|_{t_i}=\dfrac{\partial \mathcal{L}}{\partial \dot{q}_i}\delta q_i -\dfrac{\partial \mathcal{L}}{\partial \dot{q}_i}\dot{q}_i\delta t_i,
\ea
where $\delta q_i =q'(t_i')-q(t_i)$. 
The variation of the action becomes the total derivative as follows:
\ba\label{DStt}
\delta S=-\mathcal{H}\delta t_2+\mathcal{H}\delta t_1+\dfrac{\partial \mathcal{L}}{\partial \dot{q}}\delta q_2-\dfrac{\partial \mathcal{L}}{\partial \dot{q}}\delta q_1.
\ea
It shows that 
\ba
\dfrac{\partial S}{\partial t_2}=-\mathcal{H}, \quad \dfrac{\partial S}{\partial t_1}=\mathcal{H},\quad \dfrac{\partial S}{\partial q_2}=\dfrac{\partial \mathcal{L}}{\partial \dot{q_2}}=p_2,\quad  \dfrac{\partial S}{\partial q_1}=-\dfrac{\partial \mathcal{L}}{\partial \dot{q_1}}=-p_1.
\ea
Thus, the motion in which \eqref{DStt} becomes the total derivative is possible.

\section{Hamilton-Jacobi equations for small $l$}\label{appB}
We derive the Hamilton-Jacobi equation for a striped shape for $d=5,\ 6$ in this appendix. When we restrict to the small size $l$ of the subregion, we can use the analytic expression. The Hamilton Jacobi equation is given by eq. \eqref{HJs} and eq. \eqref{ENTd1}. 

We consider $d=5$ first. The turning point, $z_t$ is expanded in the small $l$ limit (see \cite{Fujita:2020qvp}), 
\be\bal 
z_t=&\dfrac{5l \Gamma \left(\frac{9}{8}\right)}{2 \sqrt{\pi } \Gamma \left(\frac{13}{8}\right)}-\dfrac{15625 l^6 \bar{a}_{\phi}\left(3 \Gamma \left(\frac{3}{4}\right) \Gamma \left(\frac{9}{8}\right)^6 \Gamma \left(\frac{13}{8}\right)-5 \Gamma \left(\frac{9}{8}\right)^7 \Gamma \left(\frac{5}{4}\right)\right)}{1536 \pi ^3 z_h^5 \Gamma \left(\frac{3}{4}\right) \Gamma \left(\frac{13}{8}\right)^7}   \\
&-\dfrac{1953125 l^9 (\bar{a}_{\phi}-1)  \Gamma \left(\frac{9}{8}\right)^9}{2304 \pi ^{9/2} z_h^8 \Gamma \left(\frac{13}{8}\right)^9}+O(l^{11}).
\eal\ee
The Hamilton-Jacobi equation becomes
\be\bal 
&\dfrac{dA_s}{V_2L_{\phi}L^4dl}= \dfrac{\sqrt{f(z_t)}}{z_t^{4}}   \\
=&\dfrac{16 \pi ^2 \Gamma \left(\frac{13}{8}\right)^4}{625 l^4 \Gamma \left(\frac{9}{8}\right)^4}-\dfrac{25 l \bar{a}_{\phi} \Gamma \left(\frac{9}{8}\right)^2 \Gamma \left(\frac{5}{4}\right)}{12 \sqrt{\pi } z_h^5 \Gamma \left(\frac{3}{4}\right) \Gamma \left(\frac{13}{8}\right)^2}
+\dfrac{15625 (\bar{a}_{\phi}-1) l^4 \Gamma \left(\frac{9}{8}\right)^4}{288 \pi ^2 z_h^8 \Gamma \left(\frac{13}{8}\right)^4}+O(l^6).
\eal\ee 
For $d=6$, $z_t$ is expanded in terms of small $l$ as follows:
\be\bal 
z_t=&\dfrac{3 l \Gamma \left(\frac{11}{10}\right)}{\sqrt{\pi } \Gamma \left(\frac{8}{5}\right)}
-\dfrac{2187 \bar{a}_{\phi} l^7 \left(7 \Gamma \left(\frac{7}{10}\right) \Gamma \left(\frac{8}{5}\right)-12 \Gamma \left(\frac{11}{10}\right) \Gamma \left(\frac{6}{5}\right)\right) \Gamma \left(\frac{11}{10}\right)^7}{70 \pi ^{7/2} z_h^6 \Gamma \left(\frac{7}{10}\right) \Gamma \left(\frac{8}{5}\right)^8}  \\
&-\dfrac{885735 (\bar{a}_{\phi}-1) l^{11} \Gamma \left(\frac{11}{10}\right)^{11}}{22 \pi ^{11/2} z_h^{10} \Gamma \left(\frac{8}{5}\right)^{11}}+O(l^{13}).
\eal\ee 
The Hamilton-Jacobi equation becomes
\ba 
&\dfrac{dA_s}{V_3L_{\phi}L^5dl}=\dfrac{\sqrt{f(z_t)}}{z_t^5}=\dfrac{\pi ^{5/2} \Gamma \left(\frac{8}{5}\right)^5}{243 l^5 \Gamma \left(\frac{11}{10}\right)^5}
-\dfrac{18 l \bar{a}_{\phi}  \Gamma \left(\frac{11}{10}\right)^2 \Gamma \left(\frac{6}{5}\right)}{7 \sqrt{\pi } z_h^6 \Gamma \left(\frac{7}{10}\right) \Gamma \left(\frac{8}{5}\right)^2} \nonumber \\
&+l^5 (\bar{a}_{\phi}-1) \dfrac{4374  \Gamma \left(\frac{11}{10}\right)^5}{11 \pi^{5/2} z_h^{10} \Gamma \left(\frac{8}{5}\right)^5}+O(l^7).
\ea


\begin{thebibliography}{999}
\parskip=-2pt
\bibitem{Cardy}
C.~Holzhey, F.~Larsen and F.~Wilczek,
``Geometric and renormalized entropy in conformal field theory,''
Nucl.\ Phys.\ B {\bf 424}, 443 (1994)
[arXiv:hep-th/9403108];
%%CITATION = HEP-TH 9403108;%%
P.~Calabrese and J.~L.~Cardy,
``Entanglement entropy and quantum field theory,''
J.\ Stat.\ Mech.\ {\bf 0406}, P002 (2004)
[arXiv:hep-th/0405152].
%%CITATION = HEP-TH 0405152;%%

\bibitem{Review}
P.~Calabrese and J.~Cardy,
``Entanglement entropy and conformal field theory,''
J.\ Phys.\ A {\bf 42} (2009) 504005
[arXiv:0905.4013 [cond-mat.stat-mech]];
%%CITATION = JPAGB,A42,504005;%%
H.~Casini and M.~Huerta,
``Entanglement entropy in free quantum field theory,''
J.\ Phys.\ A {\bf 42} (2009) 504007
[arXiv:0905.2562 [hep-th]].
%%CITATION = JPAGB,A42,504007;%%

%\cite{Vidal:2002rm}
\bibitem{Vidal:2002rm}
  G.~Vidal, J.~I.~Latorre, E.~Rico and A.~Kitaev,
  ``Entanglement in quantum critical phenomena,''
  Phys.\ Rev.\ Lett.\  {\bf 90}, 227902 (2003)
  % doi:10.1103/PhysRevLett.90.227902
  [quant-ph/0211074].
  %%CITATION = doi:10.1103/PhysRevLett.90.227902;%%

\bibitem{EE}
L.~Bombelli, R.~K.~Koul, J.~H.~Lee and R.~D.~Sorkin,
``A Quantum Source of Entropy for Black Holes,''
Phys.\ Rev.\  D {\bf 34}, 373 (1986);
%%CITATION = PHRVA,D34,373;%%
M.~Srednicki,
``Entropy and area,''
{Phys.\ Rev.\ Lett.}\ {\bf 71}, 666 (1993)
[arXiv:hep-th/9303048].
%%CITATION = HEP-TH 9303048;%%


%\cite{Ryu:2006bv}
\bibitem{Ryu:2006bv}
  S.~Ryu and T.~Takayanagi,
  ``Holographic derivation of entanglement entropy from AdS/CFT,''
  Phys.\ Rev.\ Lett.\  {\bf 96}, 181602 (2006)
  % doi:10.1103/PhysRevLett.96.181602
  [hep-th/0603001].
  %%CITATION = doi:10.1103/PhysRevLett.96.181602;%%

%\cite{Ryu:2006ef}
\bibitem{Ryu:2006ef}
  S.~Ryu and T.~Takayanagi,
  ``Aspects of Holographic Entanglement Entropy,''
  JHEP {\bf 0608}, 045 (2006)
  %doi:10.1088/1126-6708/2006/08/045
  [hep-th/0605073].
  %%CITATION = doi:10.1088/1126-6708/2006/08/045;%%

%\cite{Nishioka:2009un}
\bibitem{Nishioka:2009un}
  T.~Nishioka, S.~Ryu and T.~Takayanagi,
  ``Holographic Entanglement Entropy: An Overview,''
  J.\ Phys.\ A {\bf 42}, 504008 (2009)
  % doi:10.1088/1751-8113/42/50/504008
  [arXiv:0905.0932 [hep-th]].
  %%CITATION = doi:10.1088/1751-8113/42/50/504008;%%


%\cite{Casini:2004bw}
\bibitem{Casini:2004bw}
H.~Casini and M.~Huerta,
``A Finite entanglement entropy and the c-theorem,''
Phys. Lett. B \textbf{600}, 142-150 (2004)
%doi:10.1016/j.physletb.2004.08.072
[arXiv:hep-th/0405111 [hep-th]].
%323 citations counted in INSPIRE as of 08 Mar 2023

%\cite{Casini:2006es}
\bibitem{Casini:2006es}
  H.~Casini and M.~Huerta,
  ``A c-theorem for the entanglement entropy,''
  J.\ Phys.\ A {\bf 40}, 7031 (2007)
 % doi:10.1088/1751-8113/40/25/S57
  [cond-mat/0610375].
  %%CITATION = doi:10.1088/1751-8113/40/25/S57;%%
  %90 citations counted in INSPIRE as of 27 Apr 2020

%\cite{Nishioka:2006gr}
\bibitem{Nishioka:2006gr}
  T.~Nishioka and T.~Takayanagi,
  ``AdS Bubbles, Entropy and Closed String Tachyons,''
  JHEP {\bf 0701}, 090 (2007)
 % doi:10.1088/1126-6708/2007/01/090
  [hep-th/0611035].
  %%CITATION = doi:10.1088/1126-6708/2007/01/090;%%

%\cite{Klebanov:2007ws}
\bibitem{Klebanov:2007ws}
  I.~R.~Klebanov, D.~Kutasov and A.~Murugan,
 ``Entanglement as a probe of confinement,''
  Nucl.\ Phys.\ B {\bf 796}, 274 (2008)
 % doi:10.1016/j.nuclphysb.2007.12.017
  [arXiv:0709.2140 [hep-th]].
  %%CITATION = doi:10.1016/j.nuclphysb.2007.12.017;%%

%\cite{Buividovich:2008gq}
\bibitem{Buividovich:2008gq}
  P.~V.~Buividovich and M.~I.~Polikarpov,
  ``Entanglement entropy in gauge theories and the holographic principle for electric strings,''
  Phys.\ Lett.\ B {\bf 670}, 141 (2008)
%  doi:10.1016/j.physletb.2008.10.032
  [arXiv:0806.3376 [hep-th]].
  %%CITATION = doi:10.1016/j.physletb.2008.10.032;%%

%\cite{Dudal:2016joz}
\bibitem{Dudal:2016joz}
D.~Dudal and S.~Mahapatra,
``Confining gauge theories and holographic entanglement entropy with a magnetic field,''
JHEP \textbf{04}, 031 (2017)
%doi:10.1007/JHEP04(2017)031
[arXiv:1612.06248 [hep-th]].
%25 citations counted in INSPIRE as of 11 May 2020

%\cite{Dudal:2018ztm}
\bibitem{Dudal:2018ztm}
D.~Dudal and S.~Mahapatra,
``Interplay between the holographic QCD phase diagram and entanglement entropy,''
JHEP \textbf{07}, 120 (2018)
%doi:10.1007/JHEP07(2018)120
[arXiv:1805.02938 [hep-th]].
%14 citations counted in INSPIRE as of 11 May 2020

%\cite{Mahapatra:2019uql}
\bibitem{Mahapatra:2019uql}
S.~Mahapatra,
``Interplay between the holographic QCD phase diagram and mutual and n-partite information,''
JHEP \textbf{04}, 137 (2019)
%doi:10.1007/JHEP04(2019)137
[arXiv:1903.05927 [hep-th]].
%3 citations counted in INSPIRE as of 11 May 2020

%\cite{Jokela:2020wgs}
\bibitem{Jokela:2020wgs}
N.~Jokela and J.~G.~Subils,
``Is entanglement a probe of confinement?,''
JHEP \textbf{02}, 147 (2021)
%doi:10.1007/JHEP02(2021)147
[arXiv:2010.09392 [hep-th]].

%\cite{Albash:2012pd}
\bibitem{Albash:2012pd}
T.~Albash and C.~V.~Johnson,
``Holographic Studies of Entanglement Entropy in Superconductors,''
arXiv:1202.2605 [hep-th].
%%CITATION = ARXIV:1202.2605;%%

%\cite{Cai:2012sk}
\bibitem{Cai:2012sk}
R.~-G.~Cai, S.~He, L.~Li and Y.~-L.~Zhang,
``Holographic Entanglement Entropy in Insulator/Superconductor Transition,''
arXiv:1203.6620 [hep-th].
%%CITATION = ARXIV:1203.6620;%%

%\cite{Cai:2012nm}
\bibitem{Cai:2012nm}
  R.~G.~Cai, S.~He, L.~Li and Y.~L.~Zhang,
  ``Holographic Entanglement Entropy on P-wave Superconductor Phase Transition,''
  JHEP {\bf 1207}, 027 (2012)
 % doi:10.1007/JHEP07(2012)027
  [arXiv:1204.5962 [hep-th]].

%\cite{Arias:2012py}
\bibitem{Arias:2012py}
  R.~E.~Arias and I.~S.~Landea,
  ``Backreacting p-wave Superconductors,''
  JHEP {\bf 1301}, 157 (2013)
 % doi:10.1007/JHEP01(2013)157
  [arXiv:1210.6823 [hep-th]].
  %%CITATION = doi:10.1007/JHEP01(2013)157;%%

%\cite{Kuang:2014kha}
\bibitem{Kuang:2014kha}
  X.~M.~Kuang, E.~Papantonopoulos and B.~Wang,
  ``Entanglement Entropy as a Probe of the Proximity Effect in Holographic Superconductors,''
  JHEP {\bf 1405}, 130 (2014)
 % doi:10.1007/JHEP05(2014)130
  [arXiv:1401.5720 [hep-th]].
  %%CITATION = doi:10.1007/JHEP05(2014)130;%%

%\cite{Zangeneh:2017tub}
\bibitem{Zangeneh:2017tub}
  M.~K.~Zangeneh, Y.~C.~Ong and B.~Wang,
  ``Entanglement Entropy and Complexity for One-Dimensional Holographic Superconductors,''
  Phys.\ Lett.\ B {\bf 771}, 235 (2017)
 % doi:10.1016/j.physletb.2017.05.051
  [arXiv:1704.00557 [hep-th]].
  %%CITATION = doi:10.1016/j.physletb.2017.05.051;%%


%\cite{Das:2017gjy}
\bibitem{Das:2017gjy}
  S.~R.~Das, M.~Fujita and B.~S.~Kim,
  ``Holographic entanglement entropy of a 1 + 1 dimensional p-wave superconductor,''
  JHEP {\bf 1709}, 016 (2017)
 %  doi:10.1007/JHEP09(2017)016
  [arXiv:1705.10392 [hep-th]].

  %\cite{Baggioli:2023ynu}
\bibitem{Baggioli:2023ynu}
M.~Baggioli, Y.~Liu and X.~M.~Wu,
``Entanglement entropy as an order parameter for strongly coupled nodal line semimetals,''
JHEP \textbf{05}, 221 (2023)
%doi:10.1007/JHEP05(2023)221
[arXiv:2302.11096 [hep-th]].
%2 citations counted in INSPIRE as of 24 Aug 2023




%\cite{Myers:2012ed}
\bibitem{Myers:2012ed}
R.~C.~Myers and A.~Singh,
%``Comments on Holographic Entanglement Entropy and RG Flows,''
JHEP \textbf{04}, 122 (2012)
doi:10.1007/JHEP04(2012)122
[arXiv:1202.2068 [hep-th]].
%157 citations counted in INSPIRE as of 17 Aug 2023

%\cite{Fujita:2020qvp}
\bibitem{Fujita:2020qvp}
M.~Fujita, S.~He and Y.~Sun,
``Thermodynamical property of entanglement entropy and deconfinement phase transition,''
Phys. Rev. D \textbf{102} (2020) no.12, 126019
%doi:10.1103/PhysRevD.102.126019
[arXiv:2005.01048 [hep-th]].
%3 citations counted in INSPIRE as of 10 Nov 2021


%\cite{Liu:2012eea}
\bibitem{Liu:2012eea}
H.~Liu and M.~Mezei,
``A Refinement of entanglement entropy and the number of degrees of freedom,''
JHEP \textbf{04} (2013), 162
%doi:10.1007/JHEP04(2013)162
[arXiv:1202.2070 [hep-th]].
%229 citations counted in INSPIRE as of 10 Nov 2021

%\cite{Ghasemi:2019xrl}
\bibitem{Ghasemi:2019xrl}
M.~Ghasemi and S.~Parvizi,
``Constraints on anisotropic RG flows from holographic entanglement entropy,''
Phys. Rev. D \textbf{104}, 086028 (2021)
%doi:10.1103/PhysRevD.104.086028
[arXiv:1907.01546 [hep-th]].
%9 citations counted in INSPIRE as of 21 Sep 2023

%\cite{Ghasemi:2017pke}
\bibitem{Ghasemi:2017pke}
M.~Ghasemi and S.~Parvizi,
``Entanglement entropy of singular surfaces under relevant deformations in holography,''
JHEP \textbf{02}, 009 (2018)
%doi:10.1007/JHEP02(2018)009
[arXiv:1709.08169 [hep-th]].
%6 citations counted in INSPIRE as of 21 Sep 2023

%\cite{Polchinski:1998rq}
\bibitem{Polchinski:1998rq}
J.~Polchinski,
``String theory. Vol. 1: An introduction to the bosonic string,''
Cambridge University Press, 2007,
ISBN 978-0-511-25227-3, 978-0-521-67227-6, 978-0-521-63303-1
%doi:10.1017/CBO9780511816079
%615 citations counted in INSPIRE as of 19 Aug 2023

%\cite{Ishihara:2012jg}
\bibitem{Ishihara:2012jg}
M.~Ishihara, F.~L.~Lin and B.~Ning,
``Refined Holographic Entanglement Entropy for the AdS Solitons and AdS black Holes,''
Nucl. Phys. B \textbf{872}, 392-426 (2013)
%doi:10.1016/j.nuclphysb.2013.04.003
[arXiv:1203.6153 [hep-th]].
%22 citations counted in INSPIRE as of 09 Jan 2023

%\cite{Casini:2009sr}
\bibitem{Casini:2009sr}
H.~Casini and M.~Huerta,
``Entanglement entropy in free quantum field theory,''
J. Phys. A \textbf{42}, 504007 (2009)
doi:10.1088/1751-8113/42/50/504007
[arXiv:0905.2562 [hep-th]].
%585 citations counted in INSPIRE as of 12 Jun 2023

%\cite{Huerta:2011qi}
\bibitem{Huerta:2011qi}
M.~Huerta,
``Numerical Determination of the Entanglement Entropy for Free Fields in the Cylinder,''
Phys. Lett. B \textbf{710}, 691-696 (2012)
doi:10.1016/j.physletb.2012.03.044
[arXiv:1112.1277 [hep-th]].
%58 citations counted in INSPIRE as of 12 Jun 2023

%\cite{Solodukhin:2008dh}
\bibitem{Solodukhin:2008dh}
S.~N.~Solodukhin,
``Entanglement entropy, conformal invariance and extrinsic geometry,''
Phys. Lett. B \textbf{665}, 305-309 (2008)
%doi:10.1016/j.physletb.2008.05.071
[arXiv:0802.3117 [hep-th]].
%290 citations counted in INSPIRE as of 02 Apr 2023


%\cite{Horowitz:1998ha}
\bibitem{Horowitz:1998ha}
G.~T.~Horowitz and R.~C.~Myers,
``The AdS / CFT correspondence and a new positive energy conjecture for general relativity,''
Phys.\ Rev.\ D {\bf 59}, 026005 (1998)
% doi:10.1103/PhysRevD.59.026005
[hep-th/9808079].
%%CITATION = doi:10.1103/PhysRevD.59.026005;%%
%377 citations counted in INSPIRE as of 14 Feb 2020

%\cite{Hartnoll:2009sz}
\bibitem{Hartnoll:2009sz}
S.~A.~Hartnoll,
``Lectures on holographic methods for condensed matter physics,''
Class.\ Quant.\ Grav.\ {\bf 26}, 224002 (2009)
%doi:10.1088/0264-9381/26/22/224002
[arXiv:0903.3246 [hep-th]].
%%CITATION = doi:10.1088/0264-9381/26/22/224002;%%

%\cite{Balasubramanian:1999re}
\bibitem{Balasubramanian:1999re}
V.~Balasubramanian and P.~Kraus,
``A Stress tensor for Anti-de Sitter gravity,''
Commun.\ Math.\ Phys.\ {\bf 208}, 413 (1999)
% doi:10.1007/s002200050764
[hep-th/9902121].
%%CITATION = doi:10.1007/s002200050764;%%
%1435 citations counted in INSPIRE as of 03 Dec 2019

%\cite{Henningson:1998gx}
\bibitem{Henningson:1998gx}
M.~Henningson and K.~Skenderis,
``The Holographic Weyl anomaly,''
JHEP {\bf 9807}, 023 (1998)
[hep-th/9806087].

%\cite{de Haro:2000xn}
\bibitem{de Haro:2000xn}
S.~de Haro, S.~N.~Solodukhin and K.~Skenderis,
``Holographic reconstruction of space-time and renormalization in the AdS / CFT correspondence,''
Commun.\ Math.\ Phys.\ {\bf 217}, 595 (2001)
[hep-th/0002230].


%\cite{Allahbakhshi:2013wk}
\bibitem{Allahbakhshi:2013wk} 
  D.~Allahbakhshi and M.~Alishahiha,
  ``Probing Fractionalized Charges,''
  Adv.\ High Energy Phys.\  {\bf 2013}, 498068 (2013)
  % doi:10.1155/2013/498068
  [arXiv:1301.4815 [hep-th]].
  %%CITATION = doi:10.1155/2013/498068;%%

  %\cite{Bah:2008cj}
\bibitem{Bah:2008cj} 
  I.~Bah, L.~A.~Pando Zayas and C.~A.~Terrero-Escalante,
  ``Holographic Geometric Entropy at Finite Temperature from Black Holes in Global Anti de Sitter Spaces,''
  Int.\ J.\ Mod.\ Phys.\ A {\bf 27}, 1250048 (2012)
 % doi:10.1142/S0217751X12500480
  [arXiv:0809.2912 [hep-th]].
  %%CITATION = doi:10.1142/S0217751X12500480;%%

%\cite{Fujita:2008zv}
\bibitem{Fujita:2008zv} 
  M.~Fujita, T.~Nishioka and T.~Takayanagi,
  ``Geometric Entropy and Hagedorn/Deconfinement Transition,''
  JHEP {\bf 0809}, 016 (2008)
 % doi:10.1088/1126-6708/2008/09/016
  [arXiv:0806.3118 [hep-th]].
  %%CITATION = doi:10.1088/1126-6708/2008/09/016;%%

  %\cite{Witten:1998zw}
\bibitem{Witten:1998zw}
E.~Witten,
``Anti-de Sitter space, thermal phase transition, and confinement in gauge theories,''
Adv. Theor. Math. Phys. \textbf{2}, 505-532 (1998)
%doi:10.4310/ATMP.1998.v2.n3.a3
[arXiv:hep-th/9803131 [hep-th]].
%3632 citations counted in INSPIRE as of 21 Aug 2023

%\cite{Latorre:2004pk}
\bibitem{Latorre:2004pk}
J.~I.~Latorre, C.~A.~Lutken, E.~Rico and G.~Vidal,
``Fine grained entanglement loss along renormalization group flows,''
Phys. Rev. A \textbf{71}, 034301 (2005)
%doi:10.1103/PhysRevA.71.034301
[arXiv:quant-ph/0404120 [quant-ph]].
%67 citations counted in INSPIRE as of 08 Mar 2023

%\cite{Hirata:2006jx}
\bibitem{Hirata:2006jx}
T.~Hirata and T.~Takayanagi,
``AdS/CFT and strong subadditivity of entanglement entropy,''
JHEP \textbf{02}, 042 (2007)
%doi:10.1088/1126-6708/2007/02/042
[arXiv:hep-th/0608213 [hep-th]].
%154 citations counted in INSPIRE as of 21 Mar 2023




%\cite{Lieb:1973cp}\cite{Lieb:1973zz}
\bibitem{Lieb:1973cp}
E.~H.~Lieb and M.~B.~Ruskai,
``Proof of the strong subadditivity of quantum-mechanical entropy,''
J. Math. Phys. \textbf{14}, 1938-1941 (1973)
doi:10.1063/1.1666274
%196 citations counted in INSPIRE as of 12 Nov 2020

%\cite{Lieb:1973zz}
\bibitem{Lieb:1973zz}
E.~H.~Lieb and M.~B.~Ruskai,
``A Fundamental Property of Quantum-Mechanical Entropy,''
Phys. Rev. Lett. \textbf{30}, 434-436 (1973)
%doi:10.1103/PhysRevLett.30.434
%64 citations counted in INSPIRE as of 12 Nov 2020

%\cite{Csaki:1998qr}
\bibitem{Csaki:1998qr}
C.~Csaki, H.~Ooguri, Y.~Oz and J.~Terning,
``Glueball mass spectrum from supergravity,''
JHEP \textbf{01}, 017 (1999)
%doi:10.1088/1126-6708/1999/01/017
[arXiv:hep-th/9806021 [hep-th]].
%348 citations counted in INSPIRE as of 06 Sep 2023

%\cite{Deddo:2022wxj}
\bibitem{Deddo:2022wxj}
E.~Deddo, L.~A.~Pando Zayas and C.~F.~Uhlemann,
``Entanglement and topology in RG flows across dimensions: caps, bridges and corners,''
JHEP \textbf{04}, 018 (2023)
%doi:10.1007/JHEP04(2023)018
[arXiv:2301.00257 [hep-th]].
%1 citations counted in INSPIRE as of 21 Sep 2023









\end{thebibliography}
\end{document}